%% file: BcBoundsEta_2208.tex
\documentclass[reprint,twocolumn,superscriptaddress,showpacs,
nofootinbib,notitlepage,floatfix]{revtex4-1}

\usepackage{graphicx}
\usepackage{latexsym,amsmath,amssymb,lmodern,float,url}
\usepackage{natbib}
\usepackage{color}
\usepackage{microtype}
\usepackage{slashed}
\usepackage{multirow}
\usepackage{bm}

\newcommand{\be}{\begin{equation}}
\newcommand{\ee}{\end{equation}}
\newcommand{\bea}{\begin{eqnarray}}
\newcommand{\eea}{\end{eqnarray}}

\newcommand{\ba}{\begin{array}}
\newcommand{\ea}{\end{array}}

\newcommand{\rjp}{R(J/\psi)}
\newcommand{\rec}{R(\eta_c)}
\newcommand{\jp}{J/\psi}
\newcommand{\bc}{B_c^+}
\newcommand{\ec}{\eta_c}

\usepackage[colorlinks=true,backref=false, linktocpage=true,
citecolor=blue,urlcolor=blue,linkcolor=blue,pdfpagemode=UseOutlines]{hyperref}

\hypersetup{%
  bookmarksnumbered=true,
  pdftitle = {},
  pdfsubject = {},
  pdfauthor = {},
  pdfkeywords = {}
}

\begin{document}
\title{Model-Independent Prediction of $\rec$}
\author{Anson Berns}
\email{anson@umd.edu}
\affiliation{Montgomery Blair High School, Silver Spring, MD 20901, USA}
\author{Henry Lamm}
\email{hlamm@umd.edu}
\affiliation{Department of Physics, University of Maryland, College
Park, MD 20742, USA}
\date{August, 2018}

\begin{abstract}
We present a model-independent prediction for $\rec \! \equiv \!
\mathcal{BR} (\bc \rightarrow \ec \, \tau^+\nu_\tau)/ \mathcal{BR}
(\bc \rightarrow \ec \, \mu^+\nu_\mu)$.  This prediction is obtained from the form factors through a combination of
dispersive relations, heavy-quark relations at zero-recoil, and the
limited existing determinations from lattice QCD\@.  The resulting prediction, $R(\eta_c)=0.29(5)$, agrees with the weighted average of previous model predictions, but with reduced uncertainties.
\end{abstract}

\maketitle
\section{Introduction}

The Higgs interaction is the only source of {\it lepton universality\/} violations within the standard model, but the observation of neutrino masses implies
that at least one form of
beyond-standard model modification exist.  The ratios of semileptonic
heavy-meson decays for distinct lepton flavors are particularly sensitive to new physics, because the QCD
dynamics of the heavy-meson decays decouple from the electroweak
interaction at leading order:
\begin{equation}\label{eq:matel}
|\mathcal{M}_{\bar{b}\rightarrow\bar{c} \, \ell^+ \nu_\ell}|^2=
\frac{L_{\mu\nu}H^{\mu\nu}}{q^2-M_W^2}+\mathcal{O}(\alpha,G_F)\,.
\end{equation}
This expression implies that the ratios of semileptonic heavy-meson
decays can differ from unity at this level of precision only due to
kinematic factors, although it is possible to further remove this dependence~\cite{Colangelo:2018cnj,
Ivanov:2016qtw,Tran:2018kuv,Bhattacharya:2015ida,Bhattacharya:2016zcw,
Bhattacharya:2018kig, Jaiswal:2017rve,Cohen:2018vhw}.  Measurements from BaBar, Belle, and LHCb of the
ratios $R(D^{(*)})$ of heavy-light meson decays $B \! \rightarrow \!
D^{(*)}\ell\bar{\nu}$, with $\ell \! = \! \tau$ to $\ell \! = \!
\mu$, exhibit tension with theoretical predictions.  The HFLAV averages~\cite{Amhis:2016xyh} of
the experimental results $R(D^*) \! = \!
0.306(13)(7)$~\cite{Lees:2012xj,Lees:2013uzd,Huschle:2015rga,
Sato:2016svk,Aaij:2015yra,Hirose:2016wfn,Aaij:2017uff,Aaij:2017deq,Hirose:2017dxl}
and $R(D) \! = \!
0.407(39)(24)$~\cite{Lees:2012xj,Lees:2013uzd,Huschle:2015rga}
represent a combined 3.8$\sigma$
discrepancy~\cite{Amhis:2016xyh} from the HFLAV-suggested
Standard-Model value of $R(D^*) \! = \! 0.258(5)$~\cite{Amhis:2016xyh}
obtained by an
averaging~\cite{Bernlochner:2017jka,Bigi:2017jbd,Jaiswal:2017rve} that
utilizes experimental form factors, lattice QCD results, and
heavy-quark effective theory, and from $R(D) \! = \!
0.300(8)$~\cite{Aoki:2016frl}, which is an average of lattice QCD
results~\cite{Lattice:2015rga,Na:2015kha}, as well as a value $R(D) \!
= \! 0.299(3)$ obtained by also including experimentally
extracted form factors~\cite{Bigi:2016mdz}.  Recently, the LHCb collaboration
has measured $R(\jp) =
0.71(17)(18)$~\cite{Aaij:2017tyk} which agrees with the Standard-Model bound of $0.20\leq R(J/\psi)\leq0.39$ at 1.3$\sigma$~\cite{Cohen:2018dgz}.  In the future, it would be useful to consider the $\bar{b}c\rightarrow\bar{c}c$ analog of the $B\rightarrow D$ process, $\bc\rightarrow\ec$.  Alas, measurements of $\rec$ are substantially harder than $\rjp$ for a few reasons, foremost of which is there is no clean process like $\jp\rightarrow\mu^+\mu^-$ in which to reconstruct the $\ec$, which will result in larger backgrounds.  Additionally the transition to $\eta_c$ from excited states is poorly understood, and this further complicates extraction of signals~\cite{BHHJ}. 

Despite these present experimental difficulties, it would be valuable to have a theoretical prediction for $\rec$ from the Standard Model ready for it.  The current state of affairs, though, is limited to model-dependent calculations
(collected in Table~\ref{tab:models})~\cite{Anisimov:1998xv,Kiselev:1999sc,
Ivanov:2000aj,Kiselev:2002vz,Ivanov:2006ni,Hernandez:2006gt,
Qiao:2012vt,Wen-Fei:2013uea,Rui:2016opu,Dutta:2017xmj,
Liptaj:2017nks,Issadykov:2018myx,Tran:2018kuv}.  Although most
models' central values cluster in the range $0.25-0.35$, one notes a wide spread in their estimated
uncertainty which typically account only for parameter fitting.  We take as a reasonable estimate the weighted average of the results, $\rec=0.33(17)$.  These results rely upon
some approximations to obtain the $\bc\rightarrow\ec$ transition form factors.  Without a clear understanding
of the systematic uncertainties these assumptions introduce, the
reliability of these predictions is suspect.

\begin{figure}[ht]
 \includegraphics[width=\linewidth]{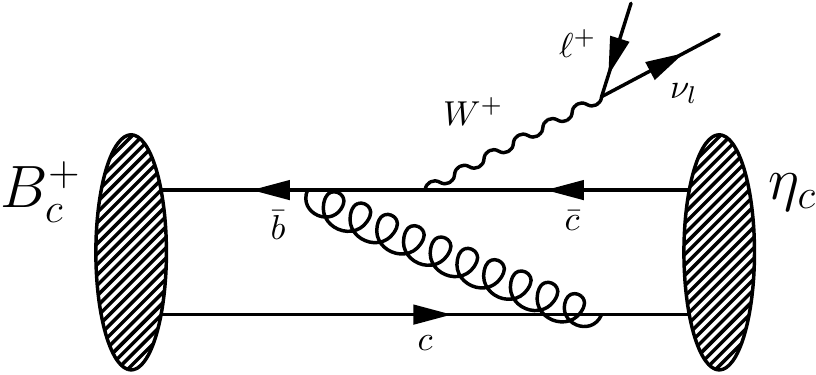}
 \caption{\label{fig:bcdecay}Schematic picture of the $\bc\rightarrow
  \ec \, \ell^+ \nu_\ell$ process.}
\end{figure}

We fill a blank space in the literature by computing a model-independent prediction,
$\rec=0.29(5)$ from the Standard Model, in
which all uncertainties are quantifiable.  In order to obtain this
result, we begin in Sec.~\ref{sec:sm} with a discussion of the $V-A$
structure of the Standard Model and the form factors.  In
Sec.~\ref{sec:hqss} we explain how heavy-quark spin symmetry can be
applied at the zero-recoil point to relate the form factors, using the
method of~\cite{Kiselev:1999sc}.  The initial lattice QCD results of
the HPQCD collaboration~\cite{Colquhoun:2016osw,*ALE} for the
transition form factors are discussed in Sec.~\ref{sec:lat}\@.  The
dispersive analysis framework utilized to constrain the form factors
as functions of momentum transfer is presented in Sec.~\ref{sec:da}\@.
The results of our analysis, as well as future projections, appear in Sec.~\ref{sec:results}, and we conclude in
Sec.~\ref{sec:con}.

After this calculation was completed, a similar calculation appeared~\cite{Murphy:2018sqg} that is in good agreement with ours.

\section{Structure of $\langle\ec|(V-A)^\mu|\bc
\rangle$}\label{sec:sm}
In the Standard Model, the factorization of Eq.~(\ref{eq:matel}) into
a leptonic and a hadronic tensor reduces the problem of calculating
$\rec$ to the computation of the hadronic matrix element $\langle\ec|(V-A)^\mu|\bc\rangle$.  Using this factorization,
the hadronic matrix element can be written in terms of two transition
form factors.  These form factors enter the matrix element in
combination with the meson masses, $M \! \equiv \! M_{\bc}$ and $m \!
\equiv \!  M_{\ec}$, and the corresponding meson momenta $P^\mu$ and
$p^\mu$.  The form
factors themselves depend only upon $t\equiv q^2 \! = \! (P-p)^2$, the squared
momentum transfer to the leptons.  The hadronic matrix element in our convention is given by $f_+(t),f_-(t)$:
\begin{align}\label{eq:hadme}
 \langle \ec(p)|(V-A)^\mu|\bc(P)\rangle=&f_+(P+p)^\mu+f_-(P-p)^\mu
\end{align}
In this work, we will exchange $f_-$ for $f_0$, which is given by
\begin{equation}
\label{eq:f0}
 f_0(t)=(M^2-m^2)f_++tf_-(t)
\end{equation}
In this convention, it can be seen that $f_+(0)=f_0(0)$, which should be applied when fitting the functions.  We further introduce two important kinematic values $t_{\pm}=(M\pm m)^2$.  This convention differs from that utilize by HPQCD for their lattice QCD results~\cite{Colquhoun:2016osw,ALE} by the mass dimension of $f_0$.  The conversion between the two is
\begin{equation}
\label{eq:convert}
 f_0=(M^2-m^2)f_0^{\rm HPQCD}
\end{equation}
Using Eq.~(\ref{eq:hadme}) or an equivalent basis, form factors are computed from models with uncontrolled approximations.  Some
models construct wave functions for the two mesons, while others
compute a perturbative distribution amplitude at $q^2 \!
\rightarrow \! 0$ and then extrapolate to larger values.  In addition, some models violate delicate form-factor
relations, such as the heavy-quark spin-symmetry relations discussed
below.  Due to these issues, it is potentially treacherous to take the all too well agreement seen between the
model predictions as a genuine estimate of the true standard model value instead of a theoretical prejudice in
modeling.

\begin{table}
\caption{\label{tab:models}Model predictions of $\rec$ classified by
method, which are abbreviated as: constituent quark model (CQM),
relativistic quark model (RCQM), QCD sum rules (QCDSR),
nonrelativistic quark model (NRQM), nonrelativistic QCD (NRQCD), and
perturbative QCD calculations (pQCD).}
\begin{center}
\begin{tabular}
{l| c c}
\hline\hline
Model & $R_{theory}$ & Year\\
\hline
CQM~\cite{Anisimov:1998xv} & 0.33  &1998   \\
QCDSR~\cite{Kiselev:1999sc}&$0.30^{+0.09}_{-0.09}$&1999\\
RCQM~\cite{Ivanov:2000aj} & 0.28  & 2000  \\
QCDSR~\cite{Kiselev:2002vz} &0.30   &2003   \\
RCQM~\cite{Ivanov:2006ni} & 0.27  &2006  \\
NRQM~\cite{Hernandez:2006gt} &$0.35^{+0.02}$   &2006   \\
NRQCD~\cite{Qiao:2012vt} &$0.30^{+0.11}_{-0.12}$ &2013   \\
pQCD~\cite{Wen-Fei:2013uea} &$0.31^{+0.12}_{-0.12}$  &2013   \\
pQCD~\cite{Rui:2016opu} &$0.6^{+0.3}_{-0.3}$ & 2016\\
pQCD~\cite{Dutta:2017xmj} &$0.30^{+0.12}_{-0.08}$ &2017\\
CQM~\cite{Liptaj:2017nks} & 0.26  &2017   \\
CQM~\cite{Issadykov:2018myx} &$0.25^{+0.08}_{-0.08}$   &2018   \\
RCQM~\cite{Tran:2018kuv}&0.26 &2018\\
\hline
Weighted Average & $0.33^{+0.17}_{-0.17}$ & --\\
\hline
\end{tabular}
\end{center}
\end{table}

The differential cross section for the semileptonic decay is
\begin{align}
\label{eq:difcof}
 \frac{d\Gamma}{dt}=\frac{G_F^2|V_{cb}|^2}{192\pi^3M^3}\frac{k}{t^{5/2}}(t-m_\ell^2)^2[4&k^2t(2t+m_\ell^2)|f_+|^2\nonumber\\&+3m_\ell^2|f_0|^2]\, .
\end{align}
where, in terms of the spatial momentum $\bm{p}$ of the $\ec$ in the
$\bc$ rest frame,
\begin{equation} \label{eq:kdef}
k \equiv M \sqrt{\frac{\bm{p}^2}{t}}
= \sqrt{\frac{(t_+-t)(t_--t)}{4t}} \, .
\end{equation}
Inspecting Eq.~(\ref{eq:difcof}), one can see that in the light
leptonic channels ($\ell=e,\mu$), the contribution from $f_0$ can be neglected, while in the $\tau$ channel it
cannot.

\section{Heavy-Quark Spin Symmetry}\label{sec:hqss}

Decays of heavy-light $Q\bar q$ systems possess enhanced symmetries in
the heavy-quark limit because operators that distinguish between heavy
quarks of different spin and flavor are suppressed by $1/m_Q$, and
their matrix elements vanish when $m_Q\rightarrow \infty$.
Consequently, all transition form factors $\langle Q'\bar q \,
|\mathcal{O}|Q \bar q\rangle$ in this limit are proportional to a
single, universal Isgur-Wise function
$\xi(w)$~\cite{Isgur:1989vq,Isgur:1989ed}, whose momentum-transfer
argument is $w$, the dot product of the initial and final heavy-light
hadron 4-velocities, $v^\mu \equiv p_M^\mu / M$ and $v'^\mu\equiv
p_m^\mu / m$, respectively:
\begin{equation} \label{eq:wdef}
w \equiv v \cdot v' = \gamma_m = \frac{E_m}{m} =
\frac{M^2 + m^2 - t}{2Mm} \, .
\end{equation}
At the zero-recoil point $t \! = \! (M \! - \! m)^2$ or $w \! = \! 1$,
the daughter hadron $m$ is at rest with respect to the parent $M$.
Indeed, one notes that $w$ equals the Lorentz factor $\gamma_m$ of $m$
in the $M$ rest frame.  The maximum value of $w$ corresponds to the
minimum momentum transfer $t$ through the virtual $W$ to the lepton
pair, which occurs when the leptons are created with minimal energy,
$t \! = \!  m_\ell^2$.

In heavy-light systems, the heavy-quark approximation corresponds to a
light quark bound in a nearly static spin-independent color field.  In
the weak decay $Q\rightarrow Q'$ between two very heavy quark flavors,
the momentum transfer $t$ to the light quark is insufficient to change
its state, and therefore the wave function of this light spectator
quark remains unaffected.  One thus concludes that $\xi(1) \! = \! 1$
at the zero-recoil (Isgur-Wise) point, yielding a absolute
normalization for the form factors.  These results are accurate up to
corrections of $\mathcal{O}(\Lambda_{\rm QCD}/m_{Q'})$.

In the decay $\bc \! \rightarrow \! \ec$, the spectator light quark is
replaced by another heavy quark, $c$ and some of these things will change.  This substitution results in a
the enhanced symmetries of the heavy-quark
limit being reduced~\cite{Jenkins:1992nb}.  First, the difference between the
heavy-quark kinetic energy operators produces energies no longer
negligible compared to those of the spectator $c$, spoiling the flavor symmetry
in heavy-heavy systems.  Furthermore, the
spectator $c$ receives a momentum transfer from the decay of $\bar{b}
\! \to \! \bar{c}$ of the same order as the momentum imparted to the
$\bar{c}$, so one cannot justify a normalization of the form factors
at the zero-recoil point based purely upon symmetry.

While the heavy-flavor symmetry is lost, the separate spin symmetries
of $\bar{b}$ and $\bar{c}$ quarks remain, with an additional spin
symmetry from the heavy spectator $c$.  Furthermore, the presence of
the heavy $c$ suggests a system that is closer to a nonrelativistic
limit than heavy-light systems.  In the $\bc \! \rightarrow \! \ec$
semileptonic decays, one further finds that
\begin{eqnarray}
  w_{\rm max} & = & w(t \! = \! m_\ell^2) =
 \frac{M^2 +m^2 - m_\ell^2}{2M m} \nonumber \\
 & \approx & 1.29 \ (\mu) , \ 1.24 \ (\tau) \, , \nonumber \\
  w_{\rm min} & = & w \left(t \! = \! (M \! - \! m)^2 \right) = 1 \, ,
\label{eq:wlimits}
\end{eqnarray}
suggesting that an expansion about the zero-recoil point may still be
reasonable.  Together, the spin symmetries imply that the two form
factors are related to a single, universal function $h$ ($\Delta$ in
Ref.~\cite{Jenkins:1992nb}), but only at the zero-recoil point, and no
symmetry-based normalization for $h$ can be
derived~\cite{Jenkins:1992nb}.

Using the trace formalism of~\cite{Falk:1990yz}, in
Ref.~\cite{Jenkins:1992nb} it was shown how to compute the relative
normalization between the four $\bar{Q}q\rightarrow \bar{Q}'q$ form
factors near the zero-recoil point [{\it i.e.}, where the spatial
momentum transfer to the spectator $q$ is $\alt {\cal O}(m_q)$].
Using these relations, $h$ was derived for a color-Coulomb potential
in Ref.~\cite{Jenkins:1992nb}.  This approximation was improved in
Ref.~\cite{Colangelo:1999zn}, where a constituent quark-model
calculation of $\mathcal{BR}(\bc \! \rightarrow \! \ec \, \ell^+
\nu_\ell)$ for $\ell=e,\mu$ but not $\tau$, was performed.  The
heavy-quark spin-symmetry relations were generalized
in~\cite{Kiselev:1999sc} to account for a momentum transfer to the
spectator quark occurring at leading order in NRQCD\@.  We reproduce
here the relation of~\cite{Kiselev:1999sc}, where the form factors
$f_+(w=1)$ and $f_0(w=1)$ are related by
\begin{equation}\label{eq:hqss}
 f_0(w=1) = \frac{8M^2(1-r)r\rho}{2(1+r)\rho+(1-r)(1-\rho)\sigma}f_+(w=1) , \ \
\end{equation}
where $r \! \equiv \! m/M$, $\rho \! \equiv \! m_{Q'} \! /m_Q$, and
$\sigma \! \equiv \! m_q/m_Q$.  These relations reproduce the
standard Isgur-Wise
result~\cite{Isgur:1989vq,Isgur:1989ed,Boyd:1997kz} when $\sigma \! =
\! 0$.  Terms that break these relations should be
$\mathcal{O}(m_c/m_b, \, \Lambda_{\rm QCD}/m_c)\approx30\%$, and we
allow conservatively for up to 50\% violations.  The heavy-quark spin
symmetry further relates the zero-recoil form factors of $\bc \!
\rightarrow \! \ec$ to those of $\bc \! \rightarrow \! \jp$, which
will be useful in the future to obtain further constraints on all six form factors.

\section{Lattice QCD Results}\label{sec:lat}

The state-of-the-art lattice QCD calculations for $\bc \! \rightarrow
\! \ec$ are limited to preliminary results from the HPQCD
Collaboration for $f_+(q^2)$ at 4 $q^2$ values and $f_0(q^2)$ at 5
$q^2$ values~\cite{Colquhoun:2016osw,*ALE}.  These results were
obtained using 2+1+1 HISQ ensembles, in which the smallest lattice
spacing is $a\approx0.09$~fm, and the $b$ quark is treated via NRQCD,
are reproduced in Fig.~\ref{fig:ffplot}.  For $q^2=t_-,0$ $f_0(q^2)$
has also been computed on coarser lattices and for lighter dynamical
$b$-quark ensembles, which are used to check the accuracy and assess
the uncertainty of the $a\approx0.09$~fm NRQCD results.  In contrast to the situation for $\rjp$, for $\rec$ both form factors have some lattice calculations, so the complications in treating unknown form factors is not required.  Instead, the dispersive relations are sufficiently constraining that a rigorous error budget smaller than our naive 20\% is the easiest way to reduce the error in $\rec$.

\section{Dispersive Relations}\label{sec:da}

In this work we fit the form factors of
$\bc\rightarrow\ec$ using analyticity and unitarity constraints on two-point Green's functions and a conformal parameterization
in the manner implemented by Boyd, Grinstein, and Lebed
(BGL)~\cite{Grinstein:2015wqa} for the decays of heavy-light hadrons.  This parameterization was extended to heavy-heavy systems in~\cite{Cohen:2018dgz} with slightly different set of free parameters to simplify the computation, which we will utilize. Here we briefly sketch the necessary components.

Consider the two-point momentum-space
Green's function $\Pi_J^{\mu \nu}$ of a vectorlike quark current,
$J^\mu \equiv \bar Q \Gamma^\mu Q' \,$.  $\Pi_J^{\mu \nu}$ can be
decomposed in different
ways~\cite{Boyd:1994tt,Boyd:1995cf,Boyd:1995tg,Boyd:1995sq,Boyd:1997kz};
in this work we decompose $\Pi_J^{\mu \nu}$ into spin-1 ($\Pi_J^T$) and
spin-0 ($\Pi_J^L$) pieces~\cite{Boyd:1997kz}:
\begin{eqnarray}
\Pi_J^{\mu\nu} (q) & \equiv & i \! \int \! d^4 x \, e^{iqx} \left< 0
\left| T J^\mu (x) J^{\dagger \nu} (0) \right| 0 \right> \nonumber \\ 
& = & \frac{1}{q^2} \left( q^\mu q^\nu - q^2
  g^{\mu\nu} \right) \Pi^T_J (q^2) + \frac{q^\mu q^\nu}{q^2} \Pi^L_J
(q^2) \, . \nonumber \\
\label{eq:twopoint}
\end{eqnarray}
From perturbative QCD (pQCD), the functions $\Pi^{L,T}_J$ require subtractions in order to be rendered finite.  The finite
dispersion relations are:
\begin{eqnarray}\label{eq:chilt}
\chi^L_J (q^2) \equiv \frac{\partial \Pi^L_J}{\partial q^2} & = &
\frac{1}{\pi} \int_0^\infty \! dt \, \frac{{\rm Im} \,
  \Pi^L_J(t)}{(t-q^2)^2} \, , \nonumber \\
\chi^T_J (q^2) \equiv \frac 1 2 \frac{\partial^2 \Pi^T_J}{\partial
  (q^2)^2} & = & \frac{1}{\pi} \int_0^\infty \! dt \, \frac{{\rm Im}
  \, \Pi^T_J(t)}{(t-q^2)^3} \, .
\end{eqnarray}
The freedom to chose a value of $q^2$ allows us to compute
$\chi (q^2)$ reliably in pQCD, far from where the two-point
function receives nonperturbative contributions.  The formal condition on $q^2$ to be in
the perturbative regime is
\begin{equation}
(m_Q + m_{Q'})\Lambda_{\rm QCD} \ll (m_Q + m_{Q'})^2 - q^2\,, 
\end{equation}
which, for $Q,Q' = c, b$, $q^2 = 0$ is clearly sufficient.  Existing
calculations of two-loop pQCD $\chi (q^2=0)$ modified by
non-perturbative vacuum contributions~\cite{Generalis:1990id,
Reinders:1980wk,Reinders:1981sy,Reinders:1984sr,Djouadi:1993ss} used
in Ref.~\cite{Boyd:1997kz} can be applied here.  An example of the
state of the art in this regard (although slightly different from the
approach used here) appears in Ref.~\cite{Bigi:2016mdz}.

The spectral functions ${\rm Im} \, \Pi_J$ can be decomposed into a
sum over the complete set of states $X$ that can couple the current
$J^\mu$ to the vacuum:
\begin{equation} \label{eq:FullPi}
{\rm Im} \, \Pi^{T,L}_J (q^2) = \frac 1 2 \sum_X (2\pi)^4 \delta^4
(q - p_X) \left| \left< 0 \left| J \right| \! X \right> \right|^2 \, .
\end{equation}
Each term in the sum is semipositive definite, thereby producing a
strict inequality for each $X$ in Eqs.~(\ref{eq:chilt}).  These
inequalities can be made stronger by including multiple $X$ at once,
as discussed in Refs.~\cite{Boyd:1997kz,Bigi:2017jbd,Jaiswal:2017rve}.  For $X$ we
include only below-threshold $\bc$ poles and a single two-body
channel, $\! \bc \! + \! \ec$, implying that our results provide very
conservative bounds.

For $\bc \! + \! \ec$, there are
lighter two-body threshold with the correct quantum numbers that must be taken into consideration.  The first physically prominent two-body production
threshold in $t$ occurs at $B \! + \! D$ (see
Table~\ref{tab:poles}).  With this fact in mind, we define a new
variable $t_{\rm bd} \! \equiv \!  (M_{B} \! \!  + \! M_{D})^2$
that corresponds to the first branch point in a given two-point
function, while the $\bc \! + \! \ec$ branch point occurs at
$t_+>t_{\rm bd}$.

With these variables, one maps the complex $t$ plane to the unit disk
in a variable $z$ (with the two sides of the branch cut forming the
unit circle $C$) using the conformal variable transformation
\begin{equation} \label{eq:zdef}
z(t;t_0) \equiv \frac{\sqrt{t_* - t} - \sqrt{t_* - t_0}}
{\sqrt{t_* - t} + \sqrt{t_* - t_0}} \, ,
\end{equation}
where $t_*$ is the branch point around which one deforms the contour,
and $t_0$ is a free parameter used to improve the convergence of
functions at small $z$.  In this mapping, $z$ is real for $t \le t_*$
and a pure phase for $t \ge t_*$.

Prior work that computed the form factors between baryons whose
threshold was above that of the lightest pair in that channel ({\it
i.e.}, $\Lambda_b\rightarrow\Lambda_c$, $\Lambda_b\rightarrow p$) took
$t_*=t_+$~\cite{Boyd:1995tg,Boyd:1997kz}, which introduces into the
region $|z|<1$ a subthreshold branch cut, meaning that the form
factors have complex nonanalyticities that cannot trivially be
removed.  To avoid this issue, we instead set $t_* \! = \! t_{bd}$,
which is possible because we are only interested in the semileptonic
decay region, $m_\ell^2\leq t\leq t_-$, which is always smaller than
$t_{bd}$.  This choice ensures that the only nonanalytic features
within the unit circle $|z| \! = \! 1$ are simple poles corresponding
to single particles $B_c^{(*)+}$, which can be removed by {\it
Blaschke factors\/} described below.  The need to avoid branch cuts
but not poles from $|z| \! < \! 1$ derives from the unique feature of
the Blaschke factors, which can remove each pole given only its
location ({\it i.e.}, mass), independent of its residue.\footnote{The
analytic significance of Blaschke factors for heavy-hadron form
factors was first noted in
Refs.~\cite{Caprini:1994fh,Caprini:1994np}.}  In contrast, correctly
accounting for a branch cut requires knowledge of both the location of
the branch point and the function along the cut.

To remove these subthreshold poles, one multiplies by $z(t;t_s)$
[using the definition of Eq.~(\ref{eq:zdef})], a Blaschke factor,
which eliminates a simple pole $t = t_s$.  Using this formalism, the
bound on each form factor $F_i(t)$ can be written as
\begin{equation} \label{eq:ff}
\frac{1}{\pi} \sum_i\int_{t_{\rm bd}}^\infty \! dt \left|
\frac{dz(t;t_0)}{dt}
\right| \left|P_i(t)  \phi_i (t;t_0) F_i(t) \right|^2 \leq 1 \, .
\end{equation}
The function $P_i(t)$ in Eq.~(\ref{eq:ff}) is a product of Blaschke
factors $z(t;t_p)$ that remove {\em dynamical\/} singularities due to
the presence of subthreshold resonant poles.  Masses corresponding to
the poles that must be removed in $\bc \! \to \! \jp$ are found in
Table~\ref{tab:poles}, organized by the channel to which each one
contributes.  These masses are from model
calculations~\cite{Eichten:1994gt}, with uncertainties that are
negligible for our purposes.

\begin{table}
\caption{\label{tab:poles}Lowest $\bc$ states needed for Blaschke
factors with $t \! < \! t_{\rm bc}$ (whose relevant two-body
threshold is indicated by ``Lowest pair'') for the $J^P$ channels of
interest.}
\begin{center}
\begin{tabular}
{l c c c}
\hline\hline
Type &  $J^P$&Lowest pair  & $M$ [GeV]\\
\hline
Vector & $1^-$&$BD$&6.337, 6.899, 7.012\\
\hline
Scalar&$0^+$&$BD$&6.700, 7.108\\
\hline
\end{tabular}
\end{center}
\end{table}

The weight function $\phi_i(t;t_0)$ is called an {\it outer
function\/} in complex analysis, and is given by
\begin{equation} \label{eq:outer}
\phi_i(t;t_0) = \tilde P_i(t) \left[ \frac{W_i(t)}{|dz(t;t_0)/dt| \,
\chi^j (q^2) (t-q^2)^{n_j}} \right]^{1/2} ,
\end{equation}
where $j \! = \! T,L$ (for which $n_j \! = \! 3,2$, respectively), the
function $\tilde P_i(t)$ is a product of factors $z(t;t_s)$ or
$\sqrt{z(t;t_s)}$ designed to remove {\em kinematical\/} singularities
at points $t = t_s<t_{\rm bc}$ from the other factors in
Eq.~(\ref{eq:ff}), and $W_i(t)$ is computable weight function
depending upon the particular form factor $F_i$.  The outer function
can be reexpressed in a general form for any particular $F_i$ as
\begin{widetext}
 \begin{align}
 \phi_i (t;t_0) & = \sqrt{\frac{n_I}{K \pi \chi}} \,
\left( \frac{t_{\rm bd} - t}{t_{\rm bd} - t_0} \right)^{\frac 1 4} \!
\left( \sqrt{t_{\rm bd} - t} + \sqrt{t_{\rm bd} - t_0} \right)
\left( t_{\rm bc} - t \right)^{\frac a 4} \! \left(
\sqrt{t_{\rm bd} - t} + \sqrt{t_{\rm bd} - t_-} \right)^{\frac b 2}
\! \left( \sqrt{t_{\rm bd} - t} + \sqrt{t_{\rm bd}} \right)^{-(c+3)}
\! , \label{eq:outer2}
\end{align}
\end{widetext}
where $n_I$ is an isospin Clebsch-Gordan factor, which is 1 for
$\bc \! \rightarrow \! \ec$.  The remaining factors are found in
Table~\ref{tab:factors}.
\begin{table}
\caption{\label{tab:factors}Inputs entering $\phi_i(t;t_0)$ in
Eq.~(\ref{eq:outer2}) for the meson form factors $F_i$.}
\begin{center}
\begin{tabular}
{l | c c c c c}
\hline\hline
$F_i$ &  $K$ & $\chi$ & $a$ & $b$ & $c$ \\
\hline
$f_+$ &48   & $\chi^T(+u)$ &3 &3&2 \\
\hline
$f_0$ &16 & $\chi^L(+u)$ &1 &1&1 \\
\hline
\end{tabular}
\end{center}
\end{table}
Transforming the dispersion-relation inequality, Eq.~(\ref{eq:ff}), into $z$-space:
\begin{equation} \label{eq:FFrelnz}
\frac{1}{2\pi i} \sum_i\oint_C \frac{dz}{z}
| \phi_i(z) P_i(z) F_i(z) |^2 \le 1 \,
,
\end{equation}
which, upon dividing out the non-analytic terms, allows the expansion
in $z$ of an analytic function:
\begin{equation} \label{eq:param}
F_i(t) = \frac{1}{|P_i(t)| \phi_i(t;t_0)} \sum_{n=0}^\infty a_{in}
z(t;t_0)^n\, .
\end{equation}
Inserting this form into Eq.~(\ref{eq:FFrelnz}), one finds that the
bound can be compactly written as a constraint on the Taylor series
coefficients:
\begin{equation} \label{eq:coeffs}
\sum_{i;n=0}^\infty a_{in}^2 \leq 1 \, .
\end{equation}
All possible functional dependences of the form factor $F_i(t)$
consistent with Eqs.~(\ref{eq:chilt}) are now incorporated into the
coefficients $a_{in}$.

It is useful to introduce a number of dimensionless parameters that
are functions of the meson masses:
\begin{align}
 r \equiv &\frac{m}{M} , \phantom{xxx}\delta \equiv \frac{m_\ell}{M}
 , \nonumber\\
 \beta \equiv &\frac{M_{B^{(*)}}}{M} , \phantom{xx}\Delta \equiv
\frac{M_{D}}{M} , \nonumber\\
 \kappa \equiv &(\beta+\Delta)^2-(1-r)^2 , \nonumber\\
 \lambda \equiv &(\beta+\Delta)^2-\delta^2 ,
\end{align}
and a parameter $N$ related to $t_0$ in Eq.~(\ref{eq:zdef}) by
\begin{equation} \label{eq:Ndefn}
N \equiv \frac{t_{\rm bd} - t_0}{t_{\rm bd} - t_-} \, .
\end{equation}
It is straightforward to compute the kinematical range for the
semileptonic process given in terms of $z$:
\begin{equation}
  z_{\rm max} = \frac{\sqrt{\lambda} - \sqrt{N\kappa}}
  {\sqrt{\lambda} + \sqrt{N\kappa}} \,,\;\; 
  z_{\rm min} - \left( \frac{\sqrt{N} - 1} {\sqrt{N} + 1}
  \right) \,. \label{eq:zlimits}
\end{equation}
The minimal (optimized) truncation error is achieved when $z_{\rm min}
= -z_{\rm max}$, which occurs when $N_{\rm opt} = \sqrt{\frac{\lambda}{\kappa}}$.
Evaluating at $N = N_{\rm opt}$, one finds
\begin{equation} \label{eq:zmaxminopt}
  z_{\rm max} = -z_{\rm min} = \frac{\lambda^{1/4} \! -\kappa^{1/4}}
  {\lambda^{1/4} \! +\kappa^{1/4}}\, ,
\end{equation}
From these expressions, we find that the semileptonic decays have
$z_{\rm max,\tau}\approx0.022$ and $z_{\rm max,\mu}\approx0.030$,
where each has a $1.3\%$ variation, depending upon whether the $BD$ or
$B^*D$ threshold is the lowest branch point, $t_{bd}$.

In the limit $t_{\rm bd}\rightarrow t_+$, one obtains
$\Delta\rightarrow r$, $\beta\rightarrow 1$, $\kappa \! \to \! 4r$,
and recovers the expressions in Ref.~\cite{Grinstein:2015wqa}.

\section{Results}\label{sec:results}
\begin{figure}[H]
 \includegraphics[width=\linewidth]{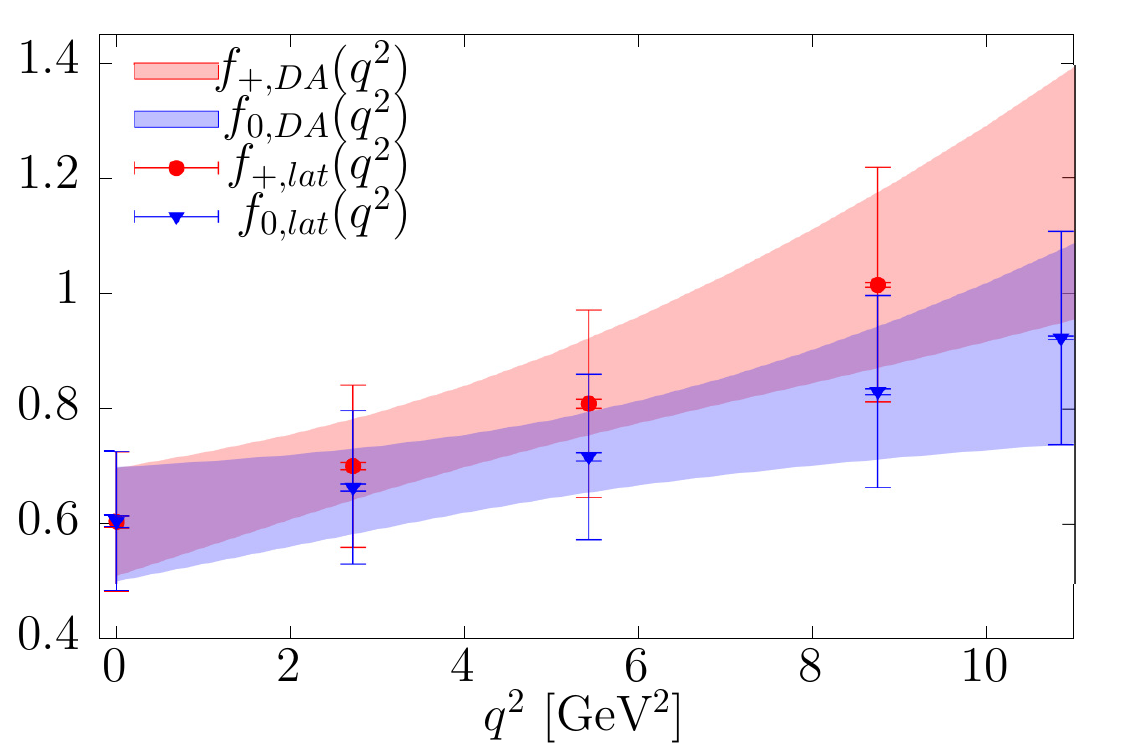}
 \caption{\label{fig:ffplot}$B^+_c \rightarrow \eta_c$ form factors $f_+(q^2)$ (red circles) and $f_0(q^2)$ (blue triangles) from the HPQCD collaboration. The interior bars represent the statistical uncertainty quoted by HPQCD. The exterior bars represent the result of including our $f_{lat} = 20\%$ systematic uncertainty. The colored bands DA (dispersive analysis) represent our one-standard-deviation ($1\sigma$) best-fit region.
}
\end{figure}

Before presenting our prediction for $R(\eta_c)$ we summarize the constraints the form factors $f_0$ and $f_+$ are required to satisfy:
\begin{itemize}
  \item The coefficients $a_n$ of each form factor are constrained by $\sum_{n} a_{n}^2 \leq 1$ from Eq.~(\ref{eq:coeffs}), in particular, for the cases $n=1,2,3$ investigated here.
  \item The form factor satisfy exactly $f_+(0)=f_0(0)$ [discussed below Eq.~(\ref{eq:f0})].
  \item Using  Eq.~(\ref{eq:hqss}), the value of $f_+(t_-)$ is required to agree with $f_0(t_-)$, which is calculated from lattice QCD, within 50\%.
\end{itemize}
Imposing these constraints, we perform our fit.  Our third assumption relating the form factors through heavy quark spin symmetry is unimposed in ~\cite{Murphy:2018sqg}, allowing us to reduce the uncertainty for $f_+(t_-)$.   Gaussian-distributed points are sampled for the form factors $f_0$ and $f_+$ whose means are given by the HPQCD results. The
combined uncertainties are given by the quadrature sum of the reported uncertainty $\delta_{lat}$
of the form-factor points and an additional systematic uncertainty, $f_{lat}$ (expressed as a percentage of the form-factor point value) that we use to estimate the uncomputed lattice uncertainties (i.e., finite-volume corrections, quark-mass dependence, discretization errors). $f_{lat}$ is taken to be 1, 5, or 20\% of the value of the form factor from the lattice.  This is a more conservative method that the $\chi^2$ procedure\cite{Murphy:2018sqg}. For our final result, we suggest using $f_{lat}$ = 20\%, while the other two values are helpful for understanding future prospects with improved lattice data. Using these sample points, we compute lines of best fit, from which we produce the coefficients $a_n$. The resulting bands of allowed form
factors are shown for $f_{lat}$ = 20\% in Fig.~\ref{fig:ffplot}, alongside the HPQCD results.

\begin{table}
\caption{\label{tab:rvalues} $R(\ec)$ as a function of the truncation power $n$ of coefficients
included from Eq.~(\ref{eq:param}) and the systematic lattice
uncertainty $f_{\rm lat}$.}
\begin{center}
\begin{tabular}
{c c c c}
\hline\hline
$f_{\rm lat}$ & $n=1$ \ & $n=2$ & $n=3$\\
\hline
1 & 0.290(4) &0.291(4) &0.290(4) \\
5 & 0.291(12) &0.291(12) &0.29(2) \\
20 &0.30(5)  &0.30(5) &0.29(5) \\
\hline
\end{tabular}
\end{center}
\end{table}

Having computed the form factors, we present predicted values for $
\rec$ as a
function of the truncation power $n=1,2,3$ in the dispersive analysis coefficients of Eq.~(\ref{eq:param}) 
and the 1, 5, 20\% systematic uncertainty $f_{lat}$ associated with the lattice data. The full
results are presented in Table \ref{tab:rvalues}, but as a conservative value, we suggest using the $n=3$, $f_{\rm lat}=20\%$ value of $\rec=0.29(5)$.  In contrast to the case of $\rjp$, we have more than three data points, and can therefore investigate the convergence more carefully.  For $f_+$, the series appears to rapidly converge such that neither $a_2$ nor $a_3$ can be distinguished from zero.  The value we obtain of $\sum_{n} a_{f_+,n}^2=0.0016(2)$ could be used to slightly strengthen bounds in future dispersive analyses in the vector channel.  For $f_0$, the typical value of $\sum_{n} a_{n}^2$ for $n=1$ is $\mathcal O(10^{-2})$, but for $n=2,3$ we find that $a_2^2 \approx 1$ despite $a_2=0.0(7)$ reflecting that while on average $a_2$ should be negligible, large fluctuations are permitted with the present uncertainties. Although the dispersive constraint is saturated in the $n\geq2$ case, the predictions for $\rec$ aren't observed to change outside of the uncertainties for increasing $n$. This confirms that while neglected higher-order terms can potentially have $a_n^2 \approx 1$, the suppression even for $z_{max} \approx 0.03$ is sufficient that the rapid convergence is still secured.
\begin{table}[H]
\caption{\label{tab:coeffs}Coefficients of $f_+$ and $f_0$ in the  expansion from Eq. \ref{eq:param} with $n=3$ and $f_{lat}=20\%$.}
\begin{center}
\begin{tabular}
{l | c c c c}
\hline\hline
& $a_0$ & $a_1$ & $a_2$ & $a_3$ \\
\hline
$f_+$ & 0.0055(6) & -0.04(3) & 0.000(10) & 0.00(6) \\
$f_0$ & 0.022(3) & -0.06(11) & 0.0(7) & 0.00(3) \\
\hline
\end{tabular}
\end{center}
\end{table}  

All model-dependent values for $\rec$ presented in Table~\ref{tab:models} comply with our result of $\rec=0.29(5)$, albeit some, e.g. the anomalously large value of $\rec=0.6(3)$ of~\cite{Rui:2016opu}, have seen their parameter space reduced.  This general agreement gives us confidence in our result.

The $B^+_c \rightarrow \eta_c$ process has sufficient $q^2$ data, with the notable exception being $f_+(t_-)$, to compute $\rec$. Following~\cite{Cohen:2018dgz}, we reanalyze our dispersive fits with a synthetic data point $f_+(t_-)=1\pm f_{\rm lat}$ to investigate it's potential constraining power. The resulting fits are found are indistinguishable from our current results within uncertainty. Therefore, the best direction for improve would be obtained by future lattice results that can fully account for the systematics we have tried to estimate.

\section{Discussion and Conclusion}\label{sec:con}

In this work we have presented a model-independent prediction of
$\rec=0.29(5)$.  While the near-term outlook for an experimental measurement of $\rec$ from  LHCb measurement is poor, near-term lattice results promises to reduce the theoretical uncertainty sufficiently to require consideration of electroweak corrections.

Even without improved lattice QCD calculations, potential areas of improvement are possible.
Experience in the heavy-light sector and the fact that the $\rjp$
bounds saturates the dispersive relations suggest that including
multiple states that appear in the dispersion relation can provides
complementary information to help constrain the form factors further, additionally one could include the lattice results for
$B\rightarrow D^{(*)}$~\cite{Lattice:2015rga,Na:2015kha,
Bailey:2014tva,Harrison:2016gup,Harrison:2017fmw,Bailey:2017xjk} and
$\Lambda_b\rightarrow \Lambda_c$~\cite{Detmold:2015aaa}.  This would allow for a global, coupled set of predictions for the semileptonic ratios.

\begin{acknowledgments}
The authors would like to thank T. Cohen, B. Hamilton, H. Jawahery, S. Lawrence, and R. Lebed for useful comments upon this paper.  H.L was supported by the U.S.\ Department of Energy under
Contract No.\ DE-FG02-93ER-40762.
\end{acknowledgments}

\bibliographystyle{apsrev4-1}
\input{BcBoundsEta_1908.bbl}
\end{document}

%% file: BcBoundsEta_1908.bbl
%

%% file: BcBoundsEta_2208.bbl
\begin{thebibliography}{66}%
\makeatletter
\providecommand \@ifxundefined [1]{%
 \@ifx{#1\undefined}
}%
\providecommand \@ifnum [1]{%
 \ifnum #1\expandafter \@firstoftwo
 \else \expandafter \@secondoftwo
 \fi
}%
\providecommand \@ifx [1]{%
 \ifx #1\expandafter \@firstoftwo
 \else \expandafter \@secondoftwo
 \fi
}%
\providecommand \natexlab [1]{#1}%
\providecommand \enquote  [1]{``#1''}%
\providecommand \bibnamefont  [1]{#1}%
\providecommand \bibfnamefont [1]{#1}%
\providecommand \citenamefont [1]{#1}%
\providecommand \href@noop [0]{\@secondoftwo}%
\providecommand \href [0]{\begingroup \@sanitize@url \@href}%
\providecommand \@href[1]{\@@startlink{#1}\@@href}%
\providecommand \@@href[1]{\endgroup#1\@@endlink}%
\providecommand \@sanitize@url [0]{\catcode `\\12\catcode `\$12\catcode
  `\&12\catcode `\#12\catcode `\^12\catcode `\_12\catcode `\%12\relax}%
\providecommand \@@startlink[1]{}%
\providecommand \@@endlink[0]{}%
\providecommand \url  [0]{\begingroup\@sanitize@url \@url }%
\providecommand \@url [1]{\endgroup\@href {#1}{\urlprefix }}%
\providecommand \urlprefix  [0]{URL }%
\providecommand \Eprint [0]{\href }%
\providecommand \doibase [0]{http://dx.doi.org/}%
\providecommand \selectlanguage [0]{\@gobble}%
\providecommand \bibinfo  [0]{\@secondoftwo}%
\providecommand \bibfield  [0]{\@secondoftwo}%
\providecommand \translation [1]{[#1]}%
\providecommand \BibitemOpen [0]{}%
\providecommand \bibitemStop [0]{}%
\providecommand \bibitemNoStop [0]{.\EOS\space}%
\providecommand \EOS [0]{\spacefactor3000\relax}%
\providecommand \BibitemShut  [1]{\csname bibitem#1\endcsname}%
\let\auto@bib@innerbib\@empty
\bibitem [{\citenamefont {Colangelo}\ and\ \citenamefont
  {De~Fazio}(2018)}]{Colangelo:2018cnj}%
  \BibitemOpen
  \bibfield  {author} {\bibinfo {author} {\bibfnamefont {P.}~\bibnamefont
  {Colangelo}}\ and\ \bibinfo {author} {\bibfnamefont {F.}~\bibnamefont
  {De~Fazio}},\ }\href {\doibase 10.1007/JHEP06(2018)082} {\bibfield  {journal}
  {\bibinfo  {journal} {JHEP}\ }\textbf {\bibinfo {volume} {06}},\ \bibinfo
  {pages} {082} (\bibinfo {year} {2018})},\ \Eprint
  {http://arxiv.org/abs/1801.10468} {arXiv:1801.10468 [hep-ph]} \BibitemShut
  {NoStop}%
\bibitem [{\citenamefont {Ivanov}\ \emph {et~al.}(2016)\citenamefont {Ivanov},
  \citenamefont {K{\" o}rner},\ and\ \citenamefont {Tran}}]{Ivanov:2016qtw}%
  \BibitemOpen
  \bibfield  {author} {\bibinfo {author} {\bibfnamefont {M.~A.}\ \bibnamefont
  {Ivanov}}, \bibinfo {author} {\bibfnamefont {J.~G.}\ \bibnamefont {K{\"
  o}rner}}, \ and\ \bibinfo {author} {\bibfnamefont {C.-T.}\ \bibnamefont
  {Tran}},\ }\href {\doibase 10.1103/PhysRevD.94.094028} {\bibfield  {journal}
  {\bibinfo  {journal} {Phys.\ Rev.}\ }\textbf {\bibinfo {volume} {D94}},\
  \bibinfo {pages} {094028} (\bibinfo {year} {2016})},\ \Eprint
  {http://arxiv.org/abs/1607.02932} {arXiv:1607.02932 [hep-ph]} \BibitemShut
  {NoStop}%
\bibitem [{\citenamefont {Tran}\ \emph {et~al.}(2018)\citenamefont {Tran},
  \citenamefont {Ivanov}, \citenamefont {K{\" o}rner},\ and\ \citenamefont
  {Santorelli}}]{Tran:2018kuv}%
  \BibitemOpen
  \bibfield  {author} {\bibinfo {author} {\bibfnamefont {C.-T.}\ \bibnamefont
  {Tran}}, \bibinfo {author} {\bibfnamefont {M.~A.}\ \bibnamefont {Ivanov}},
  \bibinfo {author} {\bibfnamefont {J.~G.}\ \bibnamefont {K{\" o}rner}}, \ and\
  \bibinfo {author} {\bibfnamefont {P.}~\bibnamefont {Santorelli}},\ }\href
  {\doibase 10.1103/PhysRevD.97.054014} {\bibfield  {journal} {\bibinfo
  {journal} {Phys.\ Rev.}\ }\textbf {\bibinfo {volume} {D97}},\ \bibinfo
  {pages} {054014} (\bibinfo {year} {2018})},\ \Eprint
  {http://arxiv.org/abs/1801.06927} {arXiv:1801.06927 [hep-ph]} \BibitemShut
  {NoStop}%
\bibitem [{\citenamefont {Bhattacharya}\ \emph {et~al.}(2016)\citenamefont
  {Bhattacharya}, \citenamefont {Nandi},\ and\ \citenamefont
  {Patra}}]{Bhattacharya:2015ida}%
  \BibitemOpen
  \bibfield  {author} {\bibinfo {author} {\bibfnamefont {S.}~\bibnamefont
  {Bhattacharya}}, \bibinfo {author} {\bibfnamefont {S.}~\bibnamefont {Nandi}},
  \ and\ \bibinfo {author} {\bibfnamefont {S.~K.}\ \bibnamefont {Patra}},\
  }\href {\doibase 10.1103/PhysRevD.93.034011} {\bibfield  {journal} {\bibinfo
  {journal} {Phys.\ Rev.}\ }\textbf {\bibinfo {volume} {D93}},\ \bibinfo
  {pages} {034011} (\bibinfo {year} {2016})},\ \Eprint
  {http://arxiv.org/abs/1509.07259} {arXiv:1509.07259 [hep-ph]} \BibitemShut
  {NoStop}%
\bibitem [{\citenamefont {Bhattacharya}\ \emph {et~al.}(2017)\citenamefont
  {Bhattacharya}, \citenamefont {Nandi},\ and\ \citenamefont
  {Patra}}]{Bhattacharya:2016zcw}%
  \BibitemOpen
  \bibfield  {author} {\bibinfo {author} {\bibfnamefont {S.}~\bibnamefont
  {Bhattacharya}}, \bibinfo {author} {\bibfnamefont {S.}~\bibnamefont {Nandi}},
  \ and\ \bibinfo {author} {\bibfnamefont {S.~K.}\ \bibnamefont {Patra}},\
  }\href {\doibase 10.1103/PhysRevD.95.075012} {\bibfield  {journal} {\bibinfo
  {journal} {Phys.\ Rev.}\ }\textbf {\bibinfo {volume} {D95}},\ \bibinfo
  {pages} {075012} (\bibinfo {year} {2017})},\ \Eprint
  {http://arxiv.org/abs/1611.04605} {arXiv:1611.04605 [hep-ph]} \BibitemShut
  {NoStop}%
\bibitem [{\citenamefont {Bhattacharya}\ \emph {et~al.}()\citenamefont
  {Bhattacharya}, \citenamefont {Nandi},\ and\ \citenamefont
  {Kumar~Patra}}]{Bhattacharya:2018kig}%
  \BibitemOpen
  \bibfield  {author} {\bibinfo {author} {\bibfnamefont {S.}~\bibnamefont
  {Bhattacharya}}, \bibinfo {author} {\bibfnamefont {S.}~\bibnamefont {Nandi}},
  \ and\ \bibinfo {author} {\bibfnamefont {S.}~\bibnamefont {Kumar~Patra}},\
  }\href@noop {} {\ }\Eprint {http://arxiv.org/abs/1805.08222}
  {arXiv:1805.08222 [hep-ph]} \BibitemShut {NoStop}%
\bibitem [{\citenamefont {Jaiswal}\ \emph {et~al.}(2017)\citenamefont
  {Jaiswal}, \citenamefont {Nandi},\ and\ \citenamefont
  {Patra}}]{Jaiswal:2017rve}%
  \BibitemOpen
  \bibfield  {author} {\bibinfo {author} {\bibfnamefont {S.}~\bibnamefont
  {Jaiswal}}, \bibinfo {author} {\bibfnamefont {S.}~\bibnamefont {Nandi}}, \
  and\ \bibinfo {author} {\bibfnamefont {S.~K.}\ \bibnamefont {Patra}},\ }\href
  {\doibase 10.1007/JHEP12(2017)060} {\bibfield  {journal} {\bibinfo  {journal}
  {JHEP}\ }\textbf {\bibinfo {volume} {12}},\ \bibinfo {pages} {060} (\bibinfo
  {year} {2017})},\ \Eprint {http://arxiv.org/abs/1707.09977} {arXiv:1707.09977
  [hep-ph]} \BibitemShut {NoStop}%
\bibitem [{\citenamefont {Cohen}\ \emph
  {et~al.}(2018{\natexlab{a}})\citenamefont {Cohen}, \citenamefont {Lamm},\
  and\ \citenamefont {Lebed}}]{Cohen:2018vhw}%
  \BibitemOpen
  \bibfield  {author} {\bibinfo {author} {\bibfnamefont {T.~D.}\ \bibnamefont
  {Cohen}}, \bibinfo {author} {\bibfnamefont {H.}~\bibnamefont {Lamm}}, \ and\
  \bibinfo {author} {\bibfnamefont {R.~F.}\ \bibnamefont {Lebed}},\ }\href@noop
  {} {\  (\bibinfo {year} {2018}{\natexlab{a}})},\ \Eprint
  {http://arxiv.org/abs/1807.00256} {arXiv:1807.00256 [hep-ph]} \BibitemShut
  {NoStop}%
\bibitem [{\citenamefont {Amhis}\ \emph {et~al.}(2017)\citenamefont {Amhis}
  \emph {et~al.}}]{Amhis:2016xyh}%
  \BibitemOpen
  \bibfield  {author} {\bibinfo {author} {\bibfnamefont {Y.}~\bibnamefont
  {Amhis}} \emph {et~al.} (\bibinfo {collaboration} {Heavy Flavor Averaging
  Group}),\ }\href {\doibase 10.1140/epjc/s10052-017-5058-4} {\bibfield
  {journal} {\bibinfo  {journal} {Eur. Phys. J.}\ }\textbf {\bibinfo {volume}
  {C77}},\ \bibinfo {pages} {895} (\bibinfo {year} {2017})},\ \bibinfo {note}
  {{updated results and plots available at
  \href{https://hflav.web.cern.ch}{{\texttt{https://hflav.web.cern.ch}}}}},\
  \Eprint {http://arxiv.org/abs/1612.07233} {arXiv:1612.07233 [hep-ex]}
  \BibitemShut {NoStop}%
\bibitem [{\citenamefont {Lees}\ \emph {et~al.}(2012)\citenamefont {Lees} \emph
  {et~al.}}]{Lees:2012xj}%
  \BibitemOpen
  \bibfield  {author} {\bibinfo {author} {\bibfnamefont {J.~P.}\ \bibnamefont
  {Lees}} \emph {et~al.} (\bibinfo {collaboration} {BaBar Collaboration}),\
  }\href {\doibase 10.1103/PhysRevLett.109.101802} {\bibfield  {journal}
  {\bibinfo  {journal} {Phys.\ Rev.\ Lett.}\ }\textbf {\bibinfo {volume}
  {109}},\ \bibinfo {pages} {101802} (\bibinfo {year} {2012})},\ \Eprint
  {http://arxiv.org/abs/1205.5442} {arXiv:1205.5442 [hep-ex]} \BibitemShut
  {NoStop}%
\bibitem [{\citenamefont {Lees}\ \emph {et~al.}(2013)\citenamefont {Lees} \emph
  {et~al.}}]{Lees:2013uzd}%
  \BibitemOpen
  \bibfield  {author} {\bibinfo {author} {\bibfnamefont {J.~P.}\ \bibnamefont
  {Lees}} \emph {et~al.} (\bibinfo {collaboration} {BaBar Collaboration}),\
  }\href {\doibase 10.1103/PhysRevD.88.072012} {\bibfield  {journal} {\bibinfo
  {journal} {Phys.\ Rev.}\ }\textbf {\bibinfo {volume} {D88}},\ \bibinfo
  {pages} {072012} (\bibinfo {year} {2013})},\ \Eprint
  {http://arxiv.org/abs/1303.0571} {arXiv:1303.0571 [hep-ex]} \BibitemShut
  {NoStop}%
\bibitem [{\citenamefont {Huschle}\ \emph {et~al.}(2015)\citenamefont {Huschle}
  \emph {et~al.}}]{Huschle:2015rga}%
  \BibitemOpen
  \bibfield  {author} {\bibinfo {author} {\bibfnamefont {M.}~\bibnamefont
  {Huschle}} \emph {et~al.} (\bibinfo {collaboration} {Belle Collaboration}),\
  }\href {\doibase 10.1103/PhysRevD.92.072014} {\bibfield  {journal} {\bibinfo
  {journal} {Phys.\ Rev.}\ }\textbf {\bibinfo {volume} {D92}},\ \bibinfo
  {pages} {072014} (\bibinfo {year} {2015})},\ \Eprint
  {http://arxiv.org/abs/1507.03233} {arXiv:1507.03233 [hep-ex]} \BibitemShut
  {NoStop}%
\bibitem [{\citenamefont {Sato}\ \emph {et~al.}(2016)\citenamefont {Sato} \emph
  {et~al.}}]{Sato:2016svk}%
  \BibitemOpen
  \bibfield  {author} {\bibinfo {author} {\bibfnamefont {Y.}~\bibnamefont
  {Sato}} \emph {et~al.} (\bibinfo {collaboration} {Belle Collaboration}),\
  }\href {\doibase 10.1103/PhysRevD.94.072007} {\bibfield  {journal} {\bibinfo
  {journal} {Phys.\ Rev.}\ }\textbf {\bibinfo {volume} {D94}},\ \bibinfo
  {pages} {072007} (\bibinfo {year} {2016})},\ \Eprint
  {http://arxiv.org/abs/1607.07923} {arXiv:1607.07923 [hep-ex]} \BibitemShut
  {NoStop}%
\bibitem [{\citenamefont {Aaij}\ \emph {et~al.}(2015)\citenamefont {Aaij} \emph
  {et~al.}}]{Aaij:2015yra}%
  \BibitemOpen
  \bibfield  {author} {\bibinfo {author} {\bibfnamefont {R.}~\bibnamefont
  {Aaij}} \emph {et~al.} (\bibinfo {collaboration} {LHCb Collaboration}),\
  }\href {\doibase 10.1103/PhysRevLett.115.159901,
  10.1103/PhysRevLett.115.111803} {\bibfield  {journal} {\bibinfo  {journal}
  {Phys.\ Rev.\ Lett.}\ }\textbf {\bibinfo {volume} {115}},\ \bibinfo {pages}
  {111803} (\bibinfo {year} {2015})},\ \bibinfo {note} {[Erratum: Phys.\ Rev.\
  Lett.\ {\bf 115}, 159901 (2015)]},\ \Eprint {http://arxiv.org/abs/1506.08614}
  {arXiv:1506.08614 [hep-ex]} \BibitemShut {NoStop}%
\bibitem [{\citenamefont {Hirose}\ \emph {et~al.}(2017)\citenamefont {Hirose}
  \emph {et~al.}}]{Hirose:2016wfn}%
  \BibitemOpen
  \bibfield  {author} {\bibinfo {author} {\bibfnamefont {S.}~\bibnamefont
  {Hirose}} \emph {et~al.} (\bibinfo {collaboration} {Belle Collaboration}),\
  }\href {\doibase 10.1103/PhysRevLett.118.211801} {\bibfield  {journal}
  {\bibinfo  {journal} {Phys.\ Rev.\ Lett.}\ }\textbf {\bibinfo {volume}
  {118}},\ \bibinfo {pages} {211801} (\bibinfo {year} {2017})},\ \Eprint
  {http://arxiv.org/abs/1612.00529} {arXiv:1612.00529 [hep-ex]} \BibitemShut
  {NoStop}%
\bibitem [{\citenamefont {Aaij}\ \emph
  {et~al.}(2018{\natexlab{a}})\citenamefont {Aaij} \emph
  {et~al.}}]{Aaij:2017uff}%
  \BibitemOpen
  \bibfield  {author} {\bibinfo {author} {\bibfnamefont {R.}~\bibnamefont
  {Aaij}} \emph {et~al.} (\bibinfo {collaboration} {LHCb}),\ }\href {\doibase
  10.1103/PhysRevLett.120.171802} {\bibfield  {journal} {\bibinfo  {journal}
  {Phys. Rev. Lett.}\ }\textbf {\bibinfo {volume} {120}},\ \bibinfo {pages}
  {171802} (\bibinfo {year} {2018}{\natexlab{a}})},\ \Eprint
  {http://arxiv.org/abs/1708.08856} {arXiv:1708.08856 [hep-ex]} \BibitemShut
  {NoStop}%
\bibitem [{\citenamefont {Aaij}\ \emph
  {et~al.}(2018{\natexlab{b}})\citenamefont {Aaij} \emph
  {et~al.}}]{Aaij:2017deq}%
  \BibitemOpen
  \bibfield  {author} {\bibinfo {author} {\bibfnamefont {R.}~\bibnamefont
  {Aaij}} \emph {et~al.} (\bibinfo {collaboration} {LHCb}),\ }\href {\doibase
  10.1103/PhysRevD.97.072013} {\bibfield  {journal} {\bibinfo  {journal} {Phys.
  Rev.}\ }\textbf {\bibinfo {volume} {D97}},\ \bibinfo {pages} {072013}
  (\bibinfo {year} {2018}{\natexlab{b}})},\ \Eprint
  {http://arxiv.org/abs/1711.02505} {arXiv:1711.02505 [hep-ex]} \BibitemShut
  {NoStop}%
\bibitem [{\citenamefont {Hirose}\ \emph {et~al.}(2018)\citenamefont {Hirose}
  \emph {et~al.}}]{Hirose:2017dxl}%
  \BibitemOpen
  \bibfield  {author} {\bibinfo {author} {\bibfnamefont {S.}~\bibnamefont
  {Hirose}} \emph {et~al.} (\bibinfo {collaboration} {Belle}),\ }\href
  {\doibase 10.1103/PhysRevD.97.012004} {\bibfield  {journal} {\bibinfo
  {journal} {Phys. Rev.}\ }\textbf {\bibinfo {volume} {D97}},\ \bibinfo {pages}
  {012004} (\bibinfo {year} {2018})},\ \Eprint
  {http://arxiv.org/abs/1709.00129} {arXiv:1709.00129 [hep-ex]} \BibitemShut
  {NoStop}%
\bibitem [{\citenamefont {Bernlochner}\ \emph {et~al.}(2017)\citenamefont
  {Bernlochner}, \citenamefont {Ligeti}, \citenamefont {Papucci},\ and\
  \citenamefont {Robinson}}]{Bernlochner:2017jka}%
  \BibitemOpen
  \bibfield  {author} {\bibinfo {author} {\bibfnamefont {F.~U.}\ \bibnamefont
  {Bernlochner}}, \bibinfo {author} {\bibfnamefont {Z.}~\bibnamefont {Ligeti}},
  \bibinfo {author} {\bibfnamefont {M.}~\bibnamefont {Papucci}}, \ and\
  \bibinfo {author} {\bibfnamefont {D.~J.}\ \bibnamefont {Robinson}},\ }\href
  {\doibase 10.1103/PhysRevD.95.115008, 10.1103/PhysRevD.97.059902} {\bibfield
  {journal} {\bibinfo  {journal} {Phys. Rev.}\ }\textbf {\bibinfo {volume}
  {D95}},\ \bibinfo {pages} {115008} (\bibinfo {year} {2017})},\ \bibinfo
  {note} {[Erratum: Phys. Rev.D97,no.5,059902(2018)]},\ \Eprint
  {http://arxiv.org/abs/1703.05330} {arXiv:1703.05330 [hep-ph]} \BibitemShut
  {NoStop}%
\bibitem [{\citenamefont {Bigi}\ \emph {et~al.}(2017)\citenamefont {Bigi},
  \citenamefont {Gambino},\ and\ \citenamefont {Schacht}}]{Bigi:2017jbd}%
  \BibitemOpen
  \bibfield  {author} {\bibinfo {author} {\bibfnamefont {D.}~\bibnamefont
  {Bigi}}, \bibinfo {author} {\bibfnamefont {P.}~\bibnamefont {Gambino}}, \
  and\ \bibinfo {author} {\bibfnamefont {S.}~\bibnamefont {Schacht}},\ }\href
  {\doibase 10.1007/JHEP11(2017)061} {\bibfield  {journal} {\bibinfo  {journal}
  {JHEP}\ }\textbf {\bibinfo {volume} {11}},\ \bibinfo {pages} {061} (\bibinfo
  {year} {2017})},\ \Eprint {http://arxiv.org/abs/1707.09509} {arXiv:1707.09509
  [hep-ph]} \BibitemShut {NoStop}%
\bibitem [{\citenamefont {Aoki}\ \emph {et~al.}(2017)\citenamefont {Aoki} \emph
  {et~al.}}]{Aoki:2016frl}%
  \BibitemOpen
  \bibfield  {author} {\bibinfo {author} {\bibfnamefont {S.}~\bibnamefont
  {Aoki}} \emph {et~al.},\ }\href {\doibase 10.1140/epjc/s10052-016-4509-7}
  {\bibfield  {journal} {\bibinfo  {journal} {Eur.\ Phys.\ J.}\ }\textbf
  {\bibinfo {volume} {C77}},\ \bibinfo {pages} {112} (\bibinfo {year}
  {2017})},\ \Eprint {http://arxiv.org/abs/1607.00299} {arXiv:1607.00299
  [hep-lat]} \BibitemShut {NoStop}%
\bibitem [{\citenamefont {Bailey}\ \emph {et~al.}(2015)\citenamefont {Bailey}
  \emph {et~al.}}]{Lattice:2015rga}%
  \BibitemOpen
  \bibfield  {author} {\bibinfo {author} {\bibfnamefont {J.~A.}\ \bibnamefont
  {Bailey}} \emph {et~al.} (\bibinfo {collaboration} {MILC Collaboration}),\
  }\href {\doibase 10.1103/PhysRevD.92.034506} {\bibfield  {journal} {\bibinfo
  {journal} {Phys.\ Rev.}\ }\textbf {\bibinfo {volume} {D92}},\ \bibinfo
  {pages} {034506} (\bibinfo {year} {2015})},\ \Eprint
  {http://arxiv.org/abs/1503.07237} {arXiv:1503.07237 [hep-lat]} \BibitemShut
  {NoStop}%
\bibitem [{\citenamefont {Na}\ \emph {et~al.}(2015)\citenamefont {Na},
  \citenamefont {Bouchard}, \citenamefont {Lepage}, \citenamefont {Monahan},\
  and\ \citenamefont {Shigemitsu}}]{Na:2015kha}%
  \BibitemOpen
  \bibfield  {author} {\bibinfo {author} {\bibfnamefont {H.}~\bibnamefont
  {Na}}, \bibinfo {author} {\bibfnamefont {C.~M.}\ \bibnamefont {Bouchard}},
  \bibinfo {author} {\bibfnamefont {G.~P.}\ \bibnamefont {Lepage}}, \bibinfo
  {author} {\bibfnamefont {C.}~\bibnamefont {Monahan}}, \ and\ \bibinfo
  {author} {\bibfnamefont {J.}~\bibnamefont {Shigemitsu}} (\bibinfo
  {collaboration} {HPQCD}),\ }\href {\doibase 10.1103/PhysRevD.93.119906,
  10.1103/PhysRevD.92.054510} {\bibfield  {journal} {\bibinfo  {journal}
  {Phys.\ Rev.}\ }\textbf {\bibinfo {volume} {D92}},\ \bibinfo {pages} {054510}
  (\bibinfo {year} {2015})},\ \bibinfo {note} {[Erratum: Phys.\ Rev.\ {\bf
  D93}, 119906 (2016)]},\ \Eprint {http://arxiv.org/abs/1505.03925}
  {arXiv:1505.03925 [hep-lat]} \BibitemShut {NoStop}%
\bibitem [{\citenamefont {Bigi}\ and\ \citenamefont
  {Gambino}(2016)}]{Bigi:2016mdz}%
  \BibitemOpen
  \bibfield  {author} {\bibinfo {author} {\bibfnamefont {D.}~\bibnamefont
  {Bigi}}\ and\ \bibinfo {author} {\bibfnamefont {P.}~\bibnamefont {Gambino}},\
  }\href {\doibase 10.1103/PhysRevD.94.094008} {\bibfield  {journal} {\bibinfo
  {journal} {Phys.\ Rev.}\ }\textbf {\bibinfo {volume} {D94}},\ \bibinfo
  {pages} {094008} (\bibinfo {year} {2016})},\ \Eprint
  {http://arxiv.org/abs/1606.08030} {arXiv:1606.08030 [hep-ph]} \BibitemShut
  {NoStop}%
\bibitem [{\citenamefont {Aaij}\ \emph
  {et~al.}(2018{\natexlab{c}})\citenamefont {Aaij} \emph
  {et~al.}}]{Aaij:2017tyk}%
  \BibitemOpen
  \bibfield  {author} {\bibinfo {author} {\bibfnamefont {R.}~\bibnamefont
  {Aaij}} \emph {et~al.} (\bibinfo {collaboration} {LHCb Collaboration}),\
  }\href {\doibase 10.1103/PhysRevLett.120.121801} {\bibfield  {journal}
  {\bibinfo  {journal} {Phys.\ Rev.\ Lett.}\ }\textbf {\bibinfo {volume}
  {120}},\ \bibinfo {pages} {121801} (\bibinfo {year} {2018}{\natexlab{c}})},\
  \Eprint {http://arxiv.org/abs/1711.05623} {arXiv:1711.05623 [hep-ex]}
  \BibitemShut {NoStop}%
\bibitem [{\citenamefont {Cohen}\ \emph
  {et~al.}(2018{\natexlab{b}})\citenamefont {Cohen}, \citenamefont {Lamm},\
  and\ \citenamefont {Lebed}}]{Cohen:2018dgz}%
  \BibitemOpen
  \bibfield  {author} {\bibinfo {author} {\bibfnamefont {T.~D.}\ \bibnamefont
  {Cohen}}, \bibinfo {author} {\bibfnamefont {H.}~\bibnamefont {Lamm}}, \ and\
  \bibinfo {author} {\bibfnamefont {R.~F.}\ \bibnamefont {Lebed}},\ }\href@noop
  {} {\  (\bibinfo {year} {2018}{\natexlab{b}})},\ \Eprint
  {http://arxiv.org/abs/1807.02730} {arXiv:1807.02730 [hep-ph]} \BibitemShut
  {NoStop}%
\bibitem [{\citenamefont {Hamilton}\ and\ \citenamefont {Jawahery}()}]{BHHJ}%
  \BibitemOpen
  \bibfield  {author} {\bibinfo {author} {\bibfnamefont {B.}~\bibnamefont
  {Hamilton}}\ and\ \bibinfo {author} {\bibfnamefont {H.}~\bibnamefont
  {Jawahery}},\ }\href@noop {} {}\bibinfo {howpublished} {personal
  communication}\BibitemShut {NoStop}%
\bibitem [{\citenamefont {Anisimov}\ \emph {et~al.}(1999)\citenamefont
  {Anisimov}, \citenamefont {Kulikov}, \citenamefont {Narodetsky},\ and\
  \citenamefont {Ter-Martirosian}}]{Anisimov:1998xv}%
  \BibitemOpen
  \bibfield  {author} {\bibinfo {author} {\bibfnamefont {A.~{\relax Yu}.}\
  \bibnamefont {Anisimov}}, \bibinfo {author} {\bibfnamefont {P.~{\relax Yu}.}\
  \bibnamefont {Kulikov}}, \bibinfo {author} {\bibfnamefont {I.~M.}\
  \bibnamefont {Narodetsky}}, \ and\ \bibinfo {author} {\bibfnamefont {K.~A.}\
  \bibnamefont {Ter-Martirosian}},\ }\href@noop {} {\bibfield  {journal}
  {\bibinfo  {journal} {Phys. Atom. Nucl.}\ }\textbf {\bibinfo {volume} {62}},\
  \bibinfo {pages} {1739} (\bibinfo {year} {1999})},\ \bibinfo {note} {[Yad.
  Fiz.62,1868(1999)]},\ \Eprint {http://arxiv.org/abs/hep-ph/9809249}
  {arXiv:hep-ph/9809249 [hep-ph]} \BibitemShut {NoStop}%
\bibitem [{\citenamefont {Kiselev}\ \emph {et~al.}(2000)\citenamefont
  {Kiselev}, \citenamefont {Likhoded},\ and\ \citenamefont
  {Onishchenko}}]{Kiselev:1999sc}%
  \BibitemOpen
  \bibfield  {author} {\bibinfo {author} {\bibfnamefont {V.~V.}\ \bibnamefont
  {Kiselev}}, \bibinfo {author} {\bibfnamefont {A.~K.}\ \bibnamefont
  {Likhoded}}, \ and\ \bibinfo {author} {\bibfnamefont {A.~I.}\ \bibnamefont
  {Onishchenko}},\ }\href {\doibase 10.1016/S0550-3213(99)00505-2} {\bibfield
  {journal} {\bibinfo  {journal} {Nucl.\ Phys.}\ }\textbf {\bibinfo {volume}
  {B569}},\ \bibinfo {pages} {473} (\bibinfo {year} {2000})},\ \Eprint
  {http://arxiv.org/abs/hep-ph/9905359} {arXiv:hep-ph/9905359 [hep-ph]}
  \BibitemShut {NoStop}%
\bibitem [{\citenamefont {Ivanov}\ \emph {et~al.}(2001)\citenamefont {Ivanov},
  \citenamefont {K{\" o}rner},\ and\ \citenamefont
  {Santorelli}}]{Ivanov:2000aj}%
  \BibitemOpen
  \bibfield  {author} {\bibinfo {author} {\bibfnamefont {M.~A.}\ \bibnamefont
  {Ivanov}}, \bibinfo {author} {\bibfnamefont {J.~G.}\ \bibnamefont {K{\"
  o}rner}}, \ and\ \bibinfo {author} {\bibfnamefont {P.}~\bibnamefont
  {Santorelli}},\ }\href {\doibase 10.1103/PhysRevD.63.074010} {\bibfield
  {journal} {\bibinfo  {journal} {Phys.\ Rev.}\ }\textbf {\bibinfo {volume}
  {D63}},\ \bibinfo {pages} {074010} (\bibinfo {year} {2001})},\ \Eprint
  {http://arxiv.org/abs/hep-ph/0007169} {arXiv:hep-ph/0007169 [hep-ph]}
  \BibitemShut {NoStop}%
\bibitem [{\citenamefont {Kiselev}()}]{Kiselev:2002vz}%
  \BibitemOpen
  \bibfield  {author} {\bibinfo {author} {\bibfnamefont {V.~V.}\ \bibnamefont
  {Kiselev}},\ }\href@noop {} {\ }\Eprint {http://arxiv.org/abs/hep-ph/0211021}
  {arXiv:hep-ph/0211021 [hep-ph]} \BibitemShut {NoStop}%
\bibitem [{\citenamefont {Ivanov}\ \emph {et~al.}(2006)\citenamefont {Ivanov},
  \citenamefont {K{\" o}rner},\ and\ \citenamefont
  {Santorelli}}]{Ivanov:2006ni}%
  \BibitemOpen
  \bibfield  {author} {\bibinfo {author} {\bibfnamefont {M.~A.}\ \bibnamefont
  {Ivanov}}, \bibinfo {author} {\bibfnamefont {J.~G.}\ \bibnamefont {K{\"
  o}rner}}, \ and\ \bibinfo {author} {\bibfnamefont {P.}~\bibnamefont
  {Santorelli}},\ }\href {\doibase 10.1103/PhysRevD.73.054024} {\bibfield
  {journal} {\bibinfo  {journal} {Phys.\ Rev.}\ }\textbf {\bibinfo {volume}
  {D73}},\ \bibinfo {pages} {054024} (\bibinfo {year} {2006})},\ \Eprint
  {http://arxiv.org/abs/hep-ph/0602050} {arXiv:hep-ph/0602050 [hep-ph]}
  \BibitemShut {NoStop}%
\bibitem [{\citenamefont {Hernandez}\ \emph {et~al.}(2006)\citenamefont
  {Hernandez}, \citenamefont {Nieves},\ and\ \citenamefont
  {Verde-Velasco}}]{Hernandez:2006gt}%
  \BibitemOpen
  \bibfield  {author} {\bibinfo {author} {\bibfnamefont {E.}~\bibnamefont
  {Hernandez}}, \bibinfo {author} {\bibfnamefont {J.}~\bibnamefont {Nieves}}, \
  and\ \bibinfo {author} {\bibfnamefont {J.~M.}\ \bibnamefont
  {Verde-Velasco}},\ }\href {\doibase 10.1103/PhysRevD.74.074008} {\bibfield
  {journal} {\bibinfo  {journal} {Phys.\ Rev.}\ }\textbf {\bibinfo {volume}
  {D74}},\ \bibinfo {pages} {074008} (\bibinfo {year} {2006})},\ \Eprint
  {http://arxiv.org/abs/hep-ph/0607150} {arXiv:hep-ph/0607150 [hep-ph]}
  \BibitemShut {NoStop}%
\bibitem [{\citenamefont {Qiao}\ and\ \citenamefont {Zhu}(2013)}]{Qiao:2012vt}%
  \BibitemOpen
  \bibfield  {author} {\bibinfo {author} {\bibfnamefont {C.-F.}\ \bibnamefont
  {Qiao}}\ and\ \bibinfo {author} {\bibfnamefont {R.-L.}\ \bibnamefont {Zhu}},\
  }\href {\doibase 10.1103/PhysRevD.87.014009} {\bibfield  {journal} {\bibinfo
  {journal} {Phys.\ Rev.}\ }\textbf {\bibinfo {volume} {D87}},\ \bibinfo
  {pages} {014009} (\bibinfo {year} {2013})},\ \Eprint
  {http://arxiv.org/abs/1208.5916} {arXiv:1208.5916 [hep-ph]} \BibitemShut
  {NoStop}%
\bibitem [{\citenamefont {Wang}\ \emph {et~al.}(2013)\citenamefont {Wang},
  \citenamefont {Fan},\ and\ \citenamefont {Xiao}}]{Wen-Fei:2013uea}%
  \BibitemOpen
  \bibfield  {author} {\bibinfo {author} {\bibfnamefont {W.-F.}\ \bibnamefont
  {Wang}}, \bibinfo {author} {\bibfnamefont {Y.-Y.}\ \bibnamefont {Fan}}, \
  and\ \bibinfo {author} {\bibfnamefont {Z.-J.}\ \bibnamefont {Xiao}},\ }\href
  {\doibase 10.1088/1674-1137/37/9/093102} {\bibfield  {journal} {\bibinfo
  {journal} {Chin.\ Phys.}\ }\textbf {\bibinfo {volume} {C37}},\ \bibinfo
  {pages} {093102} (\bibinfo {year} {2013})},\ \Eprint
  {http://arxiv.org/abs/1212.5903} {arXiv:1212.5903 [hep-ph]} \BibitemShut
  {NoStop}%
\bibitem [{\citenamefont {Rui}\ \emph {et~al.}(2016)\citenamefont {Rui},
  \citenamefont {Li}, \citenamefont {Wang},\ and\ \citenamefont
  {Xiao}}]{Rui:2016opu}%
  \BibitemOpen
  \bibfield  {author} {\bibinfo {author} {\bibfnamefont {Z.}~\bibnamefont
  {Rui}}, \bibinfo {author} {\bibfnamefont {H.}~\bibnamefont {Li}}, \bibinfo
  {author} {\bibfnamefont {G.-x.}\ \bibnamefont {Wang}}, \ and\ \bibinfo
  {author} {\bibfnamefont {Y.}~\bibnamefont {Xiao}},\ }\href {\doibase
  10.1140/epjc/s10052-016-4424-y} {\bibfield  {journal} {\bibinfo  {journal}
  {Eur.\ Phys.\ J.}\ }\textbf {\bibinfo {volume} {C76}},\ \bibinfo {pages}
  {564} (\bibinfo {year} {2016})},\ \Eprint {http://arxiv.org/abs/1602.08918}
  {arXiv:1602.08918 [hep-ph]} \BibitemShut {NoStop}%
\bibitem [{\citenamefont {Dutta}\ and\ \citenamefont
  {Bhol}(2017)}]{Dutta:2017xmj}%
  \BibitemOpen
  \bibfield  {author} {\bibinfo {author} {\bibfnamefont {R.}~\bibnamefont
  {Dutta}}\ and\ \bibinfo {author} {\bibfnamefont {A.}~\bibnamefont {Bhol}},\
  }\href {\doibase 10.1103/PhysRevD.96.076001} {\bibfield  {journal} {\bibinfo
  {journal} {Phys.\ Rev.}\ }\textbf {\bibinfo {volume} {D96}},\ \bibinfo
  {pages} {076001} (\bibinfo {year} {2017})},\ \Eprint
  {http://arxiv.org/abs/1701.08598} {arXiv:1701.08598 [hep-ph]} \BibitemShut
  {NoStop}%
\bibitem [{\citenamefont {Liptaj}\ \emph {et~al.}(2017)\citenamefont {Liptaj},
  \citenamefont {Dubnicka}, \citenamefont {Dubnickova}, \citenamefont
  {Ivanov},\ and\ \citenamefont {Issadykov}}]{Liptaj:2017nks}%
  \BibitemOpen
  \bibfield  {author} {\bibinfo {author} {\bibfnamefont {A.}~\bibnamefont
  {Liptaj}}, \bibinfo {author} {\bibfnamefont {S.}~\bibnamefont {Dubnicka}},
  \bibinfo {author} {\bibfnamefont {A.~Z.}\ \bibnamefont {Dubnickova}},
  \bibinfo {author} {\bibfnamefont {M.~A.}\ \bibnamefont {Ivanov}}, \ and\
  \bibinfo {author} {\bibfnamefont {A.}~\bibnamefont {Issadykov}},\ }\bibfield
  {booktitle} {\emph {\bibinfo {booktitle} {{Proceedings, 2017 European
  Physical Society Conference on High Energy Physics (EPS-HEP 2017): Venice,
  Italy, July 5-12, 2017}}},\ }\href {\doibase 10.22323/1.314.0667} {\bibfield
  {journal} {\bibinfo  {journal} {PoS}\ }\textbf {\bibinfo {volume}
  {EPS-HEP2017}},\ \bibinfo {pages} {667} (\bibinfo {year} {2017})}\BibitemShut
  {NoStop}%
\bibitem [{\citenamefont {Issadykov}\ and\ \citenamefont
  {Ivanov}()}]{Issadykov:2018myx}%
  \BibitemOpen
  \bibfield  {author} {\bibinfo {author} {\bibfnamefont {A.}~\bibnamefont
  {Issadykov}}\ and\ \bibinfo {author} {\bibfnamefont {M.~A.}\ \bibnamefont
  {Ivanov}},\ }\href@noop {} {\ }\Eprint {http://arxiv.org/abs/1804.00472}
  {arXiv:1804.00472 [hep-ph]} \BibitemShut {NoStop}%
\bibitem [{\citenamefont {Colquhoun}\ \emph {et~al.}(2016)\citenamefont
  {Colquhoun}, \citenamefont {Davies}, \citenamefont {Koponen}, \citenamefont
  {Lytle},\ and\ \citenamefont {McNeile}}]{Colquhoun:2016osw}%
  \BibitemOpen
  \bibfield  {author} {\bibinfo {author} {\bibfnamefont {B.}~\bibnamefont
  {Colquhoun}}, \bibinfo {author} {\bibfnamefont {C.}~\bibnamefont {Davies}},
  \bibinfo {author} {\bibfnamefont {J.}~\bibnamefont {Koponen}}, \bibinfo
  {author} {\bibfnamefont {A.}~\bibnamefont {Lytle}}, \ and\ \bibinfo {author}
  {\bibfnamefont {C.}~\bibnamefont {McNeile}} (\bibinfo {collaboration}
  {HPQCD}),\ }\bibfield  {booktitle} {\emph {\bibinfo {booktitle}
  {{Proceedings, 34th International Symposium on Lattice Field Theory (Lattice
  2016): Southampton, UK, July 24--30, 2016}}},\ }\href@noop {} {\bibfield
  {journal} {\bibinfo  {journal} {PoS}\ }\textbf {\bibinfo {volume} {LATTICE
  2016}},\ \bibinfo {pages} {281} (\bibinfo {year} {2016})},\ \Eprint
  {http://arxiv.org/abs/1611.01987} {arXiv:1611.01987 [hep-lat]} \BibitemShut
  {NoStop}%
\bibitem [{\citenamefont {Lytle}()}]{ALE}%
  \BibitemOpen
  \bibfield  {author} {\bibinfo {author} {\bibfnamefont {A.}~\bibnamefont
  {Lytle}},\ }\href@noop {} {}\bibinfo {howpublished} {personal
  communication}\BibitemShut {NoStop}%
\bibitem [{\citenamefont {Murphy}\ and\ \citenamefont
  {Soni}(2018)}]{Murphy:2018sqg}%
  \BibitemOpen
  \bibfield  {author} {\bibinfo {author} {\bibfnamefont {C.~W.}\ \bibnamefont
  {Murphy}}\ and\ \bibinfo {author} {\bibfnamefont {A.}~\bibnamefont {Soni}},\
  }\href@noop {} {\  (\bibinfo {year} {2018})},\ \Eprint
  {http://arxiv.org/abs/1808.05932} {arXiv:1808.05932 [hep-ph]} \BibitemShut
  {NoStop}%
\bibitem [{\citenamefont {Isgur}\ and\ \citenamefont
  {Wise}(1989)}]{Isgur:1989vq}%
  \BibitemOpen
  \bibfield  {author} {\bibinfo {author} {\bibfnamefont {N.}~\bibnamefont
  {Isgur}}\ and\ \bibinfo {author} {\bibfnamefont {M.~B.}\ \bibnamefont
  {Wise}},\ }\href {\doibase 10.1016/0370-2693(89)90566-2} {\bibfield
  {journal} {\bibinfo  {journal} {Phys.\ Lett.}\ }\textbf {\bibinfo {volume}
  {B232}},\ \bibinfo {pages} {113} (\bibinfo {year} {1989})}\BibitemShut
  {NoStop}%
\bibitem [{\citenamefont {Isgur}\ and\ \citenamefont
  {Wise}(1990)}]{Isgur:1989ed}%
  \BibitemOpen
  \bibfield  {author} {\bibinfo {author} {\bibfnamefont {N.}~\bibnamefont
  {Isgur}}\ and\ \bibinfo {author} {\bibfnamefont {M.~B.}\ \bibnamefont
  {Wise}},\ }\href {\doibase 10.1016/0370-2693(90)91219-2} {\bibfield
  {journal} {\bibinfo  {journal} {Phys.\ Lett.}\ }\textbf {\bibinfo {volume}
  {B237}},\ \bibinfo {pages} {527} (\bibinfo {year} {1990})}\BibitemShut
  {NoStop}%
\bibitem [{\citenamefont {Jenkins}\ \emph {et~al.}(1993)\citenamefont
  {Jenkins}, \citenamefont {Luke}, \citenamefont {Manohar},\ and\ \citenamefont
  {Savage}}]{Jenkins:1992nb}%
  \BibitemOpen
  \bibfield  {author} {\bibinfo {author} {\bibfnamefont {E.~E.}\ \bibnamefont
  {Jenkins}}, \bibinfo {author} {\bibfnamefont {M.~E.}\ \bibnamefont {Luke}},
  \bibinfo {author} {\bibfnamefont {A.~V.}\ \bibnamefont {Manohar}}, \ and\
  \bibinfo {author} {\bibfnamefont {M.~J.}\ \bibnamefont {Savage}},\ }\href
  {\doibase 10.1016/0550-3213(93)90464-Z} {\bibfield  {journal} {\bibinfo
  {journal} {Nucl.\ Phys.}\ }\textbf {\bibinfo {volume} {B390}},\ \bibinfo
  {pages} {463} (\bibinfo {year} {1993})},\ \Eprint
  {http://arxiv.org/abs/hep-ph/9204238} {arXiv:hep-ph/9204238 [hep-ph]}
  \BibitemShut {NoStop}%
\bibitem [{\citenamefont {Falk}\ \emph {et~al.}(1990)\citenamefont {Falk},
  \citenamefont {Georgi}, \citenamefont {Grinstein},\ and\ \citenamefont
  {Wise}}]{Falk:1990yz}%
  \BibitemOpen
  \bibfield  {author} {\bibinfo {author} {\bibfnamefont {A.~F.}\ \bibnamefont
  {Falk}}, \bibinfo {author} {\bibfnamefont {H.}~\bibnamefont {Georgi}},
  \bibinfo {author} {\bibfnamefont {B.}~\bibnamefont {Grinstein}}, \ and\
  \bibinfo {author} {\bibfnamefont {M.~B.}\ \bibnamefont {Wise}},\ }\href
  {\doibase 10.1016/0550-3213(90)90591-Z} {\bibfield  {journal} {\bibinfo
  {journal} {Nucl.\ Phys.}\ }\textbf {\bibinfo {volume} {B343}},\ \bibinfo
  {pages} {1} (\bibinfo {year} {1990})}\BibitemShut {NoStop}%
\bibitem [{\citenamefont {Colangelo}\ and\ \citenamefont
  {De~Fazio}(2000)}]{Colangelo:1999zn}%
  \BibitemOpen
  \bibfield  {author} {\bibinfo {author} {\bibfnamefont {P.}~\bibnamefont
  {Colangelo}}\ and\ \bibinfo {author} {\bibfnamefont {F.}~\bibnamefont
  {De~Fazio}},\ }\href {\doibase 10.1103/PhysRevD.61.034012} {\bibfield
  {journal} {\bibinfo  {journal} {Phys.\ Rev.}\ }\textbf {\bibinfo {volume}
  {D61}},\ \bibinfo {pages} {034012} (\bibinfo {year} {2000})},\ \Eprint
  {http://arxiv.org/abs/hep-ph/9909423} {arXiv:hep-ph/9909423 [hep-ph]}
  \BibitemShut {NoStop}%
\bibitem [{\citenamefont {Boyd}\ \emph {et~al.}(1997)\citenamefont {Boyd},
  \citenamefont {Grinstein},\ and\ \citenamefont {Lebed}}]{Boyd:1997kz}%
  \BibitemOpen
  \bibfield  {author} {\bibinfo {author} {\bibfnamefont {C.~G.}\ \bibnamefont
  {Boyd}}, \bibinfo {author} {\bibfnamefont {B.}~\bibnamefont {Grinstein}}, \
  and\ \bibinfo {author} {\bibfnamefont {R.~F.}\ \bibnamefont {Lebed}},\ }\href
  {\doibase 10.1103/PhysRevD.56.6895} {\bibfield  {journal} {\bibinfo
  {journal} {Phys.\ Rev.}\ }\textbf {\bibinfo {volume} {D56}},\ \bibinfo
  {pages} {6895} (\bibinfo {year} {1997})},\ \Eprint
  {http://arxiv.org/abs/hep-ph/9705252} {arXiv:hep-ph/9705252 [hep-ph]}
  \BibitemShut {NoStop}%
\bibitem [{\citenamefont {Grinstein}\ and\ \citenamefont
  {Lebed}(2015)}]{Grinstein:2015wqa}%
  \BibitemOpen
  \bibfield  {author} {\bibinfo {author} {\bibfnamefont {B.}~\bibnamefont
  {Grinstein}}\ and\ \bibinfo {author} {\bibfnamefont {R.~F.}\ \bibnamefont
  {Lebed}},\ }\href {\doibase 10.1103/PhysRevD.92.116001} {\bibfield  {journal}
  {\bibinfo  {journal} {Phys.\ Rev.}\ }\textbf {\bibinfo {volume} {D92}},\
  \bibinfo {pages} {116001} (\bibinfo {year} {2015})},\ \Eprint
  {http://arxiv.org/abs/1509.04847} {arXiv:1509.04847 [hep-ph]} \BibitemShut
  {NoStop}%
\bibitem [{\citenamefont {Boyd}\ \emph
  {et~al.}(1995{\natexlab{a}})\citenamefont {Boyd}, \citenamefont {Grinstein},\
  and\ \citenamefont {Lebed}}]{Boyd:1994tt}%
  \BibitemOpen
  \bibfield  {author} {\bibinfo {author} {\bibfnamefont {C.~G.}\ \bibnamefont
  {Boyd}}, \bibinfo {author} {\bibfnamefont {B.}~\bibnamefont {Grinstein}}, \
  and\ \bibinfo {author} {\bibfnamefont {R.~F.}\ \bibnamefont {Lebed}},\ }\href
  {\doibase 10.1103/PhysRevLett.74.4603} {\bibfield  {journal} {\bibinfo
  {journal} {Phys.\ Rev.\ Lett.}\ }\textbf {\bibinfo {volume} {74}},\ \bibinfo
  {pages} {4603} (\bibinfo {year} {1995}{\natexlab{a}})},\ \Eprint
  {http://arxiv.org/abs/hep-ph/9412324} {arXiv:hep-ph/9412324 [hep-ph]}
  \BibitemShut {NoStop}%
\bibitem [{\citenamefont {Boyd}\ \emph
  {et~al.}(1995{\natexlab{b}})\citenamefont {Boyd}, \citenamefont {Grinstein},\
  and\ \citenamefont {Lebed}}]{Boyd:1995cf}%
  \BibitemOpen
  \bibfield  {author} {\bibinfo {author} {\bibfnamefont {C.~G.}\ \bibnamefont
  {Boyd}}, \bibinfo {author} {\bibfnamefont {B.}~\bibnamefont {Grinstein}}, \
  and\ \bibinfo {author} {\bibfnamefont {R.~F.}\ \bibnamefont {Lebed}},\ }\href
  {\doibase 10.1016/0370-2693(95)00480-9} {\bibfield  {journal} {\bibinfo
  {journal} {Phys.\ Lett.}\ }\textbf {\bibinfo {volume} {B353}},\ \bibinfo
  {pages} {306} (\bibinfo {year} {1995}{\natexlab{b}})},\ \Eprint
  {http://arxiv.org/abs/hep-ph/9504235} {arXiv:hep-ph/9504235 [hep-ph]}
  \BibitemShut {NoStop}%
\bibitem [{\citenamefont {Boyd}\ and\ \citenamefont
  {Lebed}(1997)}]{Boyd:1995tg}%
  \BibitemOpen
  \bibfield  {author} {\bibinfo {author} {\bibfnamefont {C.~G.}\ \bibnamefont
  {Boyd}}\ and\ \bibinfo {author} {\bibfnamefont {R.~F.}\ \bibnamefont
  {Lebed}},\ }\href {\doibase 10.1016/S0550-3213(96)00614-1} {\bibfield
  {journal} {\bibinfo  {journal} {Nucl.\ Phys.}\ }\textbf {\bibinfo {volume}
  {B485}},\ \bibinfo {pages} {275} (\bibinfo {year} {1997})},\ \Eprint
  {http://arxiv.org/abs/hep-ph/9512363} {arXiv:hep-ph/9512363 [hep-ph]}
  \BibitemShut {NoStop}%
\bibitem [{\citenamefont {Boyd}\ \emph {et~al.}(1996)\citenamefont {Boyd},
  \citenamefont {Grinstein},\ and\ \citenamefont {Lebed}}]{Boyd:1995sq}%
  \BibitemOpen
  \bibfield  {author} {\bibinfo {author} {\bibfnamefont {C.~G.}\ \bibnamefont
  {Boyd}}, \bibinfo {author} {\bibfnamefont {B.}~\bibnamefont {Grinstein}}, \
  and\ \bibinfo {author} {\bibfnamefont {R.~F.}\ \bibnamefont {Lebed}},\ }\href
  {\doibase 10.1016/0550-3213(95)00653-2} {\bibfield  {journal} {\bibinfo
  {journal} {Nucl.\ Phys.}\ }\textbf {\bibinfo {volume} {B461}},\ \bibinfo
  {pages} {493} (\bibinfo {year} {1996})},\ \Eprint
  {http://arxiv.org/abs/hep-ph/9508211} {arXiv:hep-ph/9508211 [hep-ph]}
  \BibitemShut {NoStop}%
\bibitem [{\citenamefont {Generalis}(1990)}]{Generalis:1990id}%
  \BibitemOpen
  \bibfield  {author} {\bibinfo {author} {\bibfnamefont {S.~C.}\ \bibnamefont
  {Generalis}},\ }\href {\doibase 10.1088/0954-3899/16/6/002} {\bibfield
  {journal} {\bibinfo  {journal} {J. Phys.}\ }\textbf {\bibinfo {volume}
  {G16}},\ \bibinfo {pages} {785} (\bibinfo {year} {1990})}\BibitemShut
  {NoStop}%
\bibitem [{\citenamefont {Reinders}\ \emph {et~al.}(1980)\citenamefont
  {Reinders}, \citenamefont {Rubinstein},\ and\ \citenamefont
  {Yazaki}}]{Reinders:1980wk}%
  \BibitemOpen
  \bibfield  {author} {\bibinfo {author} {\bibfnamefont {L.~J.}\ \bibnamefont
  {Reinders}}, \bibinfo {author} {\bibfnamefont {H.~R.}\ \bibnamefont
  {Rubinstein}}, \ and\ \bibinfo {author} {\bibfnamefont {S.}~\bibnamefont
  {Yazaki}},\ }\href {\doibase 10.1016/0370-2693(80)90596-1} {\bibfield
  {journal} {\bibinfo  {journal} {Phys. Lett.}\ }\textbf {\bibinfo {volume}
  {97B}},\ \bibinfo {pages} {257} (\bibinfo {year} {1980})},\ \bibinfo {note}
  {[Erratum: Phys. Lett.100B,519(1981)]}\BibitemShut {NoStop}%
\bibitem [{\citenamefont {Reinders}\ \emph {et~al.}(1981)\citenamefont
  {Reinders}, \citenamefont {Yazaki},\ and\ \citenamefont
  {Rubinstein}}]{Reinders:1981sy}%
  \BibitemOpen
  \bibfield  {author} {\bibinfo {author} {\bibfnamefont {L.~J.}\ \bibnamefont
  {Reinders}}, \bibinfo {author} {\bibfnamefont {S.}~\bibnamefont {Yazaki}}, \
  and\ \bibinfo {author} {\bibfnamefont {H.~R.}\ \bibnamefont {Rubinstein}},\
  }\href {\doibase 10.1016/0370-2693(81)90194-5} {\bibfield  {journal}
  {\bibinfo  {journal} {Phys.\ Lett.}\ }\textbf {\bibinfo {volume} {103B}},\
  \bibinfo {pages} {63} (\bibinfo {year} {1981})}\BibitemShut {NoStop}%
\bibitem [{\citenamefont {Reinders}\ \emph {et~al.}(1985)\citenamefont
  {Reinders}, \citenamefont {Rubinstein},\ and\ \citenamefont
  {Yazaki}}]{Reinders:1984sr}%
  \BibitemOpen
  \bibfield  {author} {\bibinfo {author} {\bibfnamefont {L.~J.}\ \bibnamefont
  {Reinders}}, \bibinfo {author} {\bibfnamefont {H.}~\bibnamefont
  {Rubinstein}}, \ and\ \bibinfo {author} {\bibfnamefont {S.}~\bibnamefont
  {Yazaki}},\ }\href {\doibase 10.1016/0370-1573(85)90065-1} {\bibfield
  {journal} {\bibinfo  {journal} {Phys.\ Rept.}\ }\textbf {\bibinfo {volume}
  {127}},\ \bibinfo {pages} {1} (\bibinfo {year} {1985})}\BibitemShut {NoStop}%
\bibitem [{\citenamefont {Djouadi}\ and\ \citenamefont
  {Gambino}(1994)}]{Djouadi:1993ss}%
  \BibitemOpen
  \bibfield  {author} {\bibinfo {author} {\bibfnamefont {A.}~\bibnamefont
  {Djouadi}}\ and\ \bibinfo {author} {\bibfnamefont {P.}~\bibnamefont
  {Gambino}},\ }\href {\doibase 10.1103/PhysRevD.49.3499,
  10.1103/PhysRevD.53.4111} {\bibfield  {journal} {\bibinfo  {journal} {Phys.\
  Rev.}\ }\textbf {\bibinfo {volume} {D49}},\ \bibinfo {pages} {3499} (\bibinfo
  {year} {1994})},\ \bibinfo {note} {[Erratum: Phys.\ Rev.\ {\bf D53},4111
  (1996)]},\ \Eprint {http://arxiv.org/abs/hep-ph/9309298}
  {arXiv:hep-ph/9309298 [hep-ph]} \BibitemShut {NoStop}%
\bibitem [{\citenamefont {Caprini}(1994{\natexlab{a}})}]{Caprini:1994fh}%
  \BibitemOpen
  \bibfield  {author} {\bibinfo {author} {\bibfnamefont {I.}~\bibnamefont
  {Caprini}},\ }\href {\doibase 10.1007/BF01552631} {\bibfield  {journal}
  {\bibinfo  {journal} {Z. Phys.}\ }\textbf {\bibinfo {volume} {C61}},\
  \bibinfo {pages} {651} (\bibinfo {year} {1994}{\natexlab{a}})}\BibitemShut
  {NoStop}%
\bibitem [{\citenamefont {Caprini}(1994{\natexlab{b}})}]{Caprini:1994np}%
  \BibitemOpen
  \bibfield  {author} {\bibinfo {author} {\bibfnamefont {I.}~\bibnamefont
  {Caprini}},\ }\href {\doibase 10.1016/0370-2693(94)91153-3} {\bibfield
  {journal} {\bibinfo  {journal} {Phys.\ Lett.}\ }\textbf {\bibinfo {volume}
  {B339}},\ \bibinfo {pages} {187} (\bibinfo {year} {1994}{\natexlab{b}})},\
  \Eprint {http://arxiv.org/abs/hep-ph/9408238} {arXiv:hep-ph/9408238 [hep-ph]}
  \BibitemShut {NoStop}%
\bibitem [{\citenamefont {Eichten}\ and\ \citenamefont
  {Quigg}(1994)}]{Eichten:1994gt}%
  \BibitemOpen
  \bibfield  {author} {\bibinfo {author} {\bibfnamefont {E.~J.}\ \bibnamefont
  {Eichten}}\ and\ \bibinfo {author} {\bibfnamefont {C.}~\bibnamefont
  {Quigg}},\ }\href {\doibase 10.1103/PhysRevD.49.5845} {\bibfield  {journal}
  {\bibinfo  {journal} {Phys.\ Rev.}\ }\textbf {\bibinfo {volume} {D49}},\
  \bibinfo {pages} {5845} (\bibinfo {year} {1994})},\ \Eprint
  {http://arxiv.org/abs/hep-ph/9402210} {arXiv:hep-ph/9402210 [hep-ph]}
  \BibitemShut {NoStop}%
\bibitem [{\citenamefont {Bailey}\ \emph {et~al.}(2014)\citenamefont {Bailey}
  \emph {et~al.}}]{Bailey:2014tva}%
  \BibitemOpen
  \bibfield  {author} {\bibinfo {author} {\bibfnamefont {J.~A.}\ \bibnamefont
  {Bailey}} \emph {et~al.} (\bibinfo {collaboration} {Fermilab Lattice and MILC
  Collaborations}),\ }\href {\doibase 10.1103/PhysRevD.89.114504} {\bibfield
  {journal} {\bibinfo  {journal} {Phys.\ Rev.}\ }\textbf {\bibinfo {volume}
  {D89}},\ \bibinfo {pages} {114504} (\bibinfo {year} {2014})},\ \Eprint
  {http://arxiv.org/abs/1403.0635} {arXiv:1403.0635 [hep-lat]} \BibitemShut
  {NoStop}%
\bibitem [{\citenamefont {Harrison}\ \emph {et~al.}(2017)\citenamefont
  {Harrison}, \citenamefont {Davies},\ and\ \citenamefont
  {Wingate}}]{Harrison:2016gup}%
  \BibitemOpen
  \bibfield  {author} {\bibinfo {author} {\bibfnamefont {J.}~\bibnamefont
  {Harrison}}, \bibinfo {author} {\bibfnamefont {C.}~\bibnamefont {Davies}}, \
  and\ \bibinfo {author} {\bibfnamefont {M.}~\bibnamefont {Wingate}},\
  }\bibfield  {booktitle} {\emph {\bibinfo {booktitle} {{Proceedings, 34th
  International Symposium on Lattice Field Theory (Lattice 2016): Southampton,
  UK, July 24--30, 2016}}},\ }\href@noop {} {\bibfield  {journal} {\bibinfo
  {journal} {PoS}\ }\textbf {\bibinfo {volume} {LATTICE 2016}},\ \bibinfo
  {pages} {287} (\bibinfo {year} {2017})},\ \Eprint
  {http://arxiv.org/abs/1612.06716} {arXiv:1612.06716 [hep-lat]} \BibitemShut
  {NoStop}%
\bibitem [{\citenamefont {Harrison}\ \emph {et~al.}(2018)\citenamefont
  {Harrison}, \citenamefont {Davies},\ and\ \citenamefont
  {Wingate}}]{Harrison:2017fmw}%
  \BibitemOpen
  \bibfield  {author} {\bibinfo {author} {\bibfnamefont {J.}~\bibnamefont
  {Harrison}}, \bibinfo {author} {\bibfnamefont {C.}~\bibnamefont {Davies}}, \
  and\ \bibinfo {author} {\bibfnamefont {M.}~\bibnamefont {Wingate}} (\bibinfo
  {collaboration} {HPQCD Collaboration}),\ }\href {\doibase
  10.1103/PhysRevD.97.054502} {\bibfield  {journal} {\bibinfo  {journal}
  {Phys.\ Rev.}\ }\textbf {\bibinfo {volume} {D97}},\ \bibinfo {pages} {054502}
  (\bibinfo {year} {2018})},\ \Eprint {http://arxiv.org/abs/1711.11013}
  {arXiv:1711.11013 [hep-lat]} \BibitemShut {NoStop}%
\bibitem [{\citenamefont {Bailey}\ \emph {et~al.}(2018)\citenamefont {Bailey},
  \citenamefont {Bhattacharya}, \citenamefont {Gupta}, \citenamefont {Jang},
  \citenamefont {Lee}, \citenamefont {Leem}, \citenamefont {Park},\ and\
  \citenamefont {Yoon}}]{Bailey:2017xjk}%
  \BibitemOpen
  \bibfield  {author} {\bibinfo {author} {\bibfnamefont {J.~A.}\ \bibnamefont
  {Bailey}}, \bibinfo {author} {\bibfnamefont {T.}~\bibnamefont
  {Bhattacharya}}, \bibinfo {author} {\bibfnamefont {R.}~\bibnamefont {Gupta}},
  \bibinfo {author} {\bibfnamefont {Y.-C.}\ \bibnamefont {Jang}}, \bibinfo
  {author} {\bibfnamefont {W.}~\bibnamefont {Lee}}, \bibinfo {author}
  {\bibfnamefont {J.}~\bibnamefont {Leem}}, \bibinfo {author} {\bibfnamefont
  {S.}~\bibnamefont {Park}}, \ and\ \bibinfo {author} {\bibfnamefont
  {B.}~\bibnamefont {Yoon}} (\bibinfo {collaboration} {LANL-SWME
  Collaboration}),\ }\bibfield  {booktitle} {\emph {\bibinfo {booktitle}
  {{Proceedings, 35th International Symposium on Lattice Field Theory (Lattice
  2017): Granada, Spain, June 18--24, 2017}}},\ }\href {\doibase
  10.1051/epjconf/201817513012} {\bibfield  {journal} {\bibinfo  {journal} {EPJ
  Web Conf.}\ }\textbf {\bibinfo {volume} {175}},\ \bibinfo {pages} {13012}
  (\bibinfo {year} {2018})},\ \Eprint {http://arxiv.org/abs/1711.01786}
  {arXiv:1711.01786 [hep-lat]} \BibitemShut {NoStop}%
\bibitem [{\citenamefont {Detmold}\ \emph {et~al.}(2015)\citenamefont
  {Detmold}, \citenamefont {Lehner},\ and\ \citenamefont
  {Meinel}}]{Detmold:2015aaa}%
  \BibitemOpen
  \bibfield  {author} {\bibinfo {author} {\bibfnamefont {W.}~\bibnamefont
  {Detmold}}, \bibinfo {author} {\bibfnamefont {C.}~\bibnamefont {Lehner}}, \
  and\ \bibinfo {author} {\bibfnamefont {S.}~\bibnamefont {Meinel}},\ }\href
  {\doibase 10.1103/PhysRevD.92.034503} {\bibfield  {journal} {\bibinfo
  {journal} {Phys.\ Rev.}\ }\textbf {\bibinfo {volume} {D92}},\ \bibinfo
  {pages} {034503} (\bibinfo {year} {2015})},\ \Eprint
  {http://arxiv.org/abs/1503.01421} {arXiv:1503.01421 [hep-lat]} \BibitemShut
  {NoStop}%
\end{thebibliography}
